\documentclass[a4paper,UKenglish,cleveref, autoref, thm-restate, numberwithinsect]{lipics-v2021}

\hideLIPIcs  

\usepackage{mathtools}
\usepackage{verbatim}
\usepackage{tikz}
\usepackage{longtable,booktabs,array}

\usetikzlibrary{positioning, decorations.pathreplacing, arrows.meta, calc}

\title{Near-Optimal In-Place Sorting} 

\author{Alex Z. Xu\footnote{Corresponding author}}{Independent Researcher, Hong Kong}{xuzaxstk@gmail.com}{https://orcid.org/0009-0002-8458-2716}{}

\author{Stephen J. Chick}{Independent Researcher, Commerce City, CO, USA}{chickstephen0@gmail.com}{}{}

\authorrunning{Alex Z. Xu and Stephen J. Chick} 

\Copyright{Alex Z. Xu and Stephen J. Chick} 



\category{} 

\relatedversion{} 

\acknowledgements{}

\nolinenumbers 

\EventEditors{John Q. Open and Joan R. Access}
\EventNoEds{2}
\EventLongTitle{42nd Conference on Very Important Topics (CVIT 2016)}
\EventShortTitle{CVIT 2016}
\EventAcronym{CVIT}
\EventYear{2016}
\EventDate{December 24--27, 2016}
\EventLocation{Little Whinging, United Kingdom}
\EventLogo{}
\SeriesVolume{42}
\ArticleNo{23}

\begin{document}

\maketitle

\begin{abstract}
We present an in-place comparison-based sorting algorithm that sorts an array of $n$ elements using $n\lg n + O(n)$ comparisons with exponentially high probability and always $O(n)$ moves. This serves as the first in-place sorting algorithm to match the fundamental $n\lg n+\Omega(n)$ lower bound on total operations (i.e. sum of comparisons and moves). For the worst-case, we present an algorithm that makes $n\lg n + O(n\lg^{(t)}n)$ comparisons and $O(tn)$ data moves, where $t$ is an integer parameter satisfying $2 \leq t \leq \lg^{*}n - 1$ and $\lg^{(t)}n$ denotes the $t$-time iterated logarithm, improving over the previous upper bound of $n\lg n + O(n\lg\lg n)$ comparisons and $O(n)$ moves when using constant $t>2$. When $t\approx \lg^* n$, this yields a worst-case upper bound on total operations of $n\lg n+O(n \lg^* n)$. We thus achieve the ultimate goal of efficient in-place sorting via randomization whilst narrowing the gap to this goal in the worst-case. This advance involves a novel ordered set structure that supports searches in an optimal $\lg n + O(1)$ comparisons for $n$ elements. Note that both algorithms are unstable.
\end{abstract}

\clearpage
\section{Introduction}\label{introduction}
Sorting is one of the most fundamental problems in computer science,
carrying great practical and theoretical importance. For a comparison-based sort on a data set of \(n\) elements, it is well known that the
information-theoretic lower bound is
\(\lg(n!) \approx n\lg n - 1.4427n\) comparisons, which also holds for the average case, and at least
\(\left\lfloor 1.5n \right\rfloor\) moves are required to permute the input \cite{MR96}. Summing gives a lower bound on total operations (i.e. sum of comparisons and moves) of $\approx n\lg n+0.0573n$. Note that all logarithms throughout this paper are base 2.

In-place sorting is essential for large datasets in memory-constrained environments. A strict in-place algorithm uses only a single extra element storage location aside from the input array to permute elements and \(O(1)\) auxiliary words of size at least \(\lg n\) bits for tracking.

Simultaneously minimizing comparisons, moves, and auxiliary workspace represents a classic frontier in the complexity of sorting, dating back to foundational limits established by Munro and Raman, i.e. to achieve minimal comparisons under \(O(n)\) moves and in-place constraints. Currently, the tightest upper bound (for both
average and worst-case behavior) is
\(n\lg n + O(n\lg{\lg n})\) comparisons and
\((34 + \varepsilon)n\) moves \cite{GG11}.

Whether the \(O( n\lg{\lg n} )\) term could be reduced was left as an open problem. In this paper we resolve this open problem by presenting two variations of our in-place algorithm: a
randomized and deterministic approach. Both variations are unstable.
For any input, the randomized approach performs \(n\lg n + O(n)\) comparisons with
exponentially high probability (over the algorithm's random choices) and always \(O(n)\) moves. This probabilistically achieves the ultimate goal of in-place sorting with minimal extra comparisons and moves (see \cite{GG11, MR92}), and matches the total operation lower bound of $n\lg n+0.0573n$ up to an additive linear term.
The deterministic approach makes
\(n\lg n + O( n\lg^{(t)}n )\) comparisons and \(O(tn)\) moves
in the worst case, where \(t\) is an integer parameter satisfying
\(2 \leq t \leq \lg^{*}n - 1\), thereby narrowing the gap to this
ultimate goal: using constant $t>2$ yields a definitive improvement over \cite{GG11}. Total operations are minimized when $t \approx \lg^* n$: comparisons reach the optimal $n \lg n + O(n)$, whilst the move count degenerates slightly to $O(n \lg^* n)$.

\subsection{Technical Overview}
In \hyperref[sorting-with-additional-memory]{Section 2}, we detail the two-level structure reminiscent of y-fast tries in \cite{Wil83} that is used to facilitate sorting assuming the existence of additional memory:
\begin{itemize}
\item The upper level is an ordered set containing $s_\#$ \textit{active representatives} that supports searches in $\lg s_\# + O(1)$ comparisons and insertions in $O(\lg^3 s_\#)$ time (see \hyperref[structure-of-segment-representatives]{Section 2.2}).
\item Each active representative corresponds to lower-level \textit{segment}, which is a block of size $s=\Theta(\lg^4 s_\#)$.
\end{itemize}
The upper level's optimal search is used to route inserted elements towards segments. Elements are temporarily stored in a cache line (see Sections \hyperref[inserting-into-the-structure]{2.4} and \hyperref[structure-of-cache-lines]{2.6}) within an active representative before they are flushed into segments in batches. Only when a segment becomes full and is split into two segments each containing half the elements, is a new active representative inserted for the new segment, ensuring costly upper level insertions are sparse.

Although the upper level maintains the sorted order of segments, elements in segments are kept unsorted until extraction, where they are sorted by the procedure in \hyperref[sorting-small-subsequences]{Section 2.3}.

Lastly, in \hyperref[sorting-in-place]{Section 3} we demonstrate a method to carve out the additional memory directly from the input array, thus achieving truly in-place sorting.

The key improvements over \cite{GG11} are: the new ordered set, two optimized cache structures, and optimal segment sorting. Each change shaves off an independent $O(n \lg \lg n)$ term in comparisons, thus all three are required simultaneously to yield an improved bound.

\section{Sorting with Additional
Memory}\label{sorting-with-additional-memory}
Before considering the full in-place algorithm, we address a
simpler problem. We are given:
\begin{itemize}
\item A contiguous block \(A\) of exactly \(m\) elements, called \emph{active elements}.
\item A contiguous \emph{buffer memory} \(B\) of at least \(3m - 1\) \emph{buffer elements}. Buffer elements can be distinguished from active elements using $1$ comparison. Formally, there exists a function $\text{active}(x)$ that makes $1$ comparison then returns a Boolean \texttt{true} if $x$ is an active element and \texttt{false} otherwise (i.e. $x$ is a buffer element).
\item A bit vector \(V\) containing at least \(m/\lg^{2}m\) bits, all initialized to \texttt{false}. The bit vector supports exclusively single-bit operations: $V.\text{read}(i)$ costing $1$ comparison which returns a boolean value indicating the value of the $i$th bit and $V.\text{write}(i)$ which flips the $i$th bit costing $3$ moves.
\item \(O(1)\) auxiliary words of width \(\geq\lg m\) bits and a single empty location, i.e. the \emph{hole}; from the original in-place sorting setting.
\end{itemize}
The objective is to sort the $m$ active elements in $A$, making use of $B$ and $V$ as additional workspaces. When the program terminates, $B$ must be some permutation of its original elements, and $V$ must be reset to its initial blank state. Note that $B$ and $V$ serve as temporary mathematical abstractions of in-place mechanisms that will be fully detailed below in Section~3. If required, see \hyperref[appendix-e-notation-glossary]{[App. E]} for a complete notation glossary.

\subsection{Structure of the
Memory}\label{structure-of-the-memory}
In this section we detail the high-level layout of our provided additional workspaces.

\textbf{The buffer memory} \(B\) consists of three main sections arranged sequentially: the \textit{frame memory} \(F\),  \textit{cache memory} \(G\) and \textit{segment memory} \(S\), retaining the frame-cache-segments organization established in \cite{GG11}. \textbf{The bit vector} $V$ contains the \textit{pointer memory} $P$ and the \textit{cache data} $I$ arranged sequentially.

\textbf{The segment memory} $S$ forms the lower level of the structure, serving as the primary storage of active elements. The segment memory consists of $s_\#+1$ segments $S_0\dots S_{s_\#}$, each a contiguous block of
\begin{equation}
s = 2\left\lceil \lg m \right\rceil^{3}g - 1\
\end{equation}
\noindent
elements, where $g=\Theta(\lg m)$ is defined later in (4), giving us $s=\Theta(\lg^4 m)$. Physically in $S$, the segments are not necessarily stored sequentially in the order \(S_{0}\ldots S_{s_{\#}}\); they can be arranged in any permutation with $S_0$ at the front. The \textit{segment invariant} mandates that each segment contains between $\lfloor s/2\rfloor$ and $s$ active elements placed contiguously at the front.

Segments are dynamically allocated from a reserved zone of $2m$ elements in $B$, divided into $\lfloor 2m/s \rfloor$ blocks of $s$ elements $B_0\dots B_{\lfloor 2m/s \rfloor-1}$ (any remaining elements at the end are excluded). Initially, there is only one segment $S_0$. To allocate a new segment, we increment $s_\#$ and return the next available block in the reserved zone ($B_{s_\#}$). Immediately after allocation, $\lfloor s/2 \rfloor$ active elements are moved into the new segment to fulfill the segment invariant.

Every segment $S_i$ has a corresponding \textbf{active representative} $R'_i$ in the upper level to coordinate routing into segments. These $s_\#$  active representatives $R'_1\dots R'_{s_\#}$ form a subset of $r_\#$ pre-allocated \emph{representatives} denoted by $R_1\dots R_{r_\#}$. Any remaining non-active representatives are classified as \emph{buffer representatives}. Each representative $R_i$ is a compound data type formed by $f_i$, $P_i$, $G_i$ and $I_i$ (defined shortly).

Specifically, the first component $f_i$ is a \emph{frame element}, and together $f_1 \dots f_{r_\#}$ constitute the \textbf{frame memory} $F$. $F$ consists of \(s_{\#}\) active elements interleaved with buffer elements; $f_i$ is active iff $R_i$ is active. The \emph{active frame elements} form the sequence \(f_{1}'\ldots f_{s_{\#}}'\) obtained by scanning \(f_{1}\ldots f_{r_{\#}}\) ignoring buffer elements. This sequence is maintained as a sorted sequence by the structure in \hyperref[structure-of-segment-representatives]{Section 2.2}. The active frame elements define the \textit{ordering invariant} that maintains relative sorted order across segments: for any active element $x \in S_i$, $f'_{i} \leq x \leq f'_{i+1}$ (logical sentinels $f'_0=-\infty$, $f'_{s_\#+1}=\infty$).

By the segment invariant, there are at least $\lfloor s/2 \rfloor$ active elements in each of the $s_\#+1$ segments, and also $s_\#$ active elements reside in the frame. Because the total number of active elements cannot exceed $m$, we conclude that $s_\#$ is at most
 \begin{equation}
s_{\#} \leq \left\lfloor \left( m - \left\lfloor s/2 \right\rfloor \right)/(\left\lfloor s/2 \right\rfloor + 1) \right\rfloor \leq \left\lfloor 2m/s \right\rfloor - 1
\end{equation}

\noindent
Consequently, the size of segment memory is bounded by:
 \begin{equation}
|S| = \left( s_{\#} + 1 \right)s \leq \ 2m
\end{equation}

\noindent
Thus, the segment memory cannot surpass the reserved zone of size $2m$.

Due to the arbitrary arrangement of the segments in the segment memory, we cannot easily locate the segment $S_i$ corresponding with $R'_i$. Thus, we introduce the \textbf{pointer memory} $P$ in $V$, consisting of the \textit{segment pointers} $P_1\dots P_{r_\#}$. Let the \textit{active segment pointers} be $P'_1\dots P'_{s_\#}$, corresponding with the active representatives $R'_1\dots R'_{s_\#}$. Each $P_i$ is a block of $\left\lfloor \lg \lfloor 2m/s \rfloor \right\rfloor+1$ bits, sufficient to address any of the $\lfloor 2m/s \rfloor$ blocks in the reserved zone by encoding an integer in binary: if $P'_i=k$, then $S_i$ occupies the block $B_k$. For all segment pointers corresponding with a buffer representative, $P_i=0$ representing a $\mathrm{NIL}$ pointer as there is no corresponding segment.

However, dereferencing a segment pointer would require $\left\lfloor \lg \lfloor 2m/s \rfloor \right\rfloor+1$ reads in $V$ using $O(\lg m)$ comparisons. To minimize overhead, we utilize one pointer dereference to operate upon a batch of $g$ elements, reducing the per-element cost of dereferencing pointers to $O(\lg m)/g=O(1)$ (see (4) below). Batch operations are facilitated by the \textbf{cache memory} \(G\) using its \(r_{\#}\) \emph{cache lines} \(G_{1}\ldots G_{r_{\#}}\), each a block of $2g$ elements operating as a specialized structure used to accumulate $g$ active elements, where
\begin{equation}
\lg m \leq g \leq 2\lg m
\end{equation}
\noindent
For now, assume that a cache line \(G_{i}\) is a black box that supports
the operations \(G_{i}.\text{insert}(x)\) and \(G_{i}.\text{flush}(x)\).
\(G_{i}.\text{insert}(x)\) exchanges the active element at \(x\) with a buffer
element in \(G_{i}\), then returns a Boolean indicating whether
\(G_{i}\) has become full. \(G_{i}.\text{flush}(x)\) swaps the active elements
in \(G_{i}\) with the contiguous block of buffer elements beginning at
\(x\). The cost of cache operations will be accounted for in \emph{cache
inserts} (calls of \(\text{insert}\)) and \emph{cache flushes} (calls of
\(\text{flush}\)). The exact implementation of cache lines is specified in
\hyperref[structure-of-cache-lines]{Section 2.6}.

Initially, all cache lines consist of only buffer elements. The \textit{active cache line} $G'_i$ corresponding with $R'_i$ temporarily stores active elements belonging in $S_i$, waiting until it accumulates $g$ active elements so they are flushed into $S_i$ as a batch. Thus, the ordering invariant that applies to $S_i$ must also apply to $G'_i$.

To support cache lines in their operations, the \textbf{cache data} \(I\) in $V$ stores \(r_{\#}\) \emph{cache integers} \(I_{1}\ldots I_{r_{\#}}\), each a block of
\(\left\lfloor \lg g \right\rfloor + 1\) bits, used to store integers
from \(0\) to \(g\). Each cache line \(G_{i}\) may use its corresponding
cache integer \(I_i\) to assist its operations.

To maintain a 1:1 correspondence between active representatives and segments,
 after a new segment is allocated, a corresponding active representative is constructed from a buffer representative \(R_{i}\) and an active element \(x\) promoted to the frame. This requires \(O( \lg m ){}\) moves: swap \(x\) with \(f_{i}\) (which makes $R_i$ active), then assign \(P_{i} \coloneqq s_{\#}\) so it points
to the \(s_{\#}\)th block in the reserved zone, i.e. the newly allocated segment.

\noindent
Testing whether a representative \(R_{i}\) is active requires a
single comparison by \(\text{active}(f_{i})\). The rank of
\(R_{i}\) is defined by its frame element, meaning \(R_{i} < R_{j}\) iff
\(f_{i} < f_{j}\). To compare an active element \(x\) and \(R_{i}\),
compare \(x\) and \(f_{i}\) directly. Exchanging two
representatives \(R_{i}\) and \(R_{j}\) is done by exchanging each
component individually: \(f_{i}\) with \(f_{j}\), \(G_{i}\) with
\(G_{j}\), \(P_{i}\) with \(P_{j}\), and \(I_i\) with \(I_j\); the
process requires \(O( \lg m )\) comparisons and
\(O( \lg m )\) moves.

\(S_{0}\) is unique in the sense that it has no corresponding active
representative and is handled separately by the auxiliary pointers $l_0$
to the next buffer position in \(S_{0}\) and $r_0$ to the end of $S_0$. \looseness=-1

\begin{figure}
\centering
\begin{tikzpicture}[
    >=Straight Barb,
    dim/.style={
        {Bar[scale=1.2]Stealth[scale=1.2]}-{Stealth[scale=1.2]Bar[scale=1.2]}, 
        thick, 
        shorten <= 1.5pt, 
        shorten >= 1.5pt
    },
    box/.style={minimum height=1.2cm, inner sep=0pt, align=center},
    font=\Large,
    scale=0.7, transform shape
]

\node[box, minimum width=1cm] (fi) {$f_i$};
\node[box, minimum width=4cm, right=0pt of fi] (F) {$F$};

\node[box, minimum width=1.5cm, right=0pt of F] (Gi) {$G_i$};
\node[box, minimum width=5cm, right=0pt of Gi] (G) {$G$};

\node[box, minimum width=2cm, right=0pt of G] (Si) {$S_i$};
\node[box, minimum width=6cm, right=0pt of Si] (S) {$S$};

\draw[thick] (fi.north west) rectangle (F.south east);
\draw[thick] (Gi.north west) rectangle (G.south east);
\draw[thick] (Si.north west) rectangle (S.south east);

\draw[thick, dashed] (fi.north east) -- (fi.south east);
\draw[thick, dashed] (Gi.north east) -- (Gi.south east);
\draw[thick, dashed] (Si.north east) -- (Si.south east);

\draw[dim] ([yshift=0.3cm]fi.north west) -- ([yshift=0.3cm]F.north east)
    node[midway, above=3pt] {$r_\#$ elements};

\draw[dim] ([yshift=0.3cm]Gi.north west) -- ([yshift=0.3cm]G.north east)
    node[midway, above=3pt] {$r_\#$ cache lines};

\draw[dim] ([yshift=0.3cm]Si.north west) -- ([yshift=0.3cm]S.north east)
    node[midway, above=3pt] {$s_\#+1$ segments};

\path (fi.north west) -- (S.north east) 
    node[midway, above=1.3cm, font=\Large] {Buffer Memory $B$ ($3m-1$ buffer elements)};

\draw[<->, thick] ([yshift=-0.3cm]Gi.south west) -- ([yshift=-0.3cm]Gi.south east)
    node[midway, below=3pt] {$2g$};

\draw[<->, thick] ([yshift=-0.3cm]Si.south west) -- ([yshift=-0.3cm]Si.south east)
    node[midway, below=3pt] {$s$};

\node[box, minimum width=3.2cm, below=3.6cm of fi.south west, anchor=north west, xshift=2.5cm] (Pi) {$P_i$};
\node[box, minimum width=4.3cm, right=0pt of Pi] (VP) {$P$};

\node[box, minimum width=2.4cm, right=0pt of VP] (Vi) {$I_i$};
\node[box, minimum width=4.6cm, right=0pt of Vi] (VG) {$I$};

\draw[thick] (Pi.north west) rectangle (VP.south east);
\draw[thick] (Vi.north west) rectangle (VG.south east);

\draw[thick, dashed] (Pi.north east) -- (Pi.south east);
\draw[thick, dashed] (Vi.north east) -- (Vi.south east);

\draw[dim] ([yshift=0.3cm]Pi.north west) -- ([yshift=0.3cm]VP.north east)
    node[midway, above=3pt] {$r_\#$ pointers};

\draw[dim] ([yshift=0.3cm]Vi.north west) -- ([yshift=0.3cm]VG.north east)
    node[midway, above=3pt] {$r_\#$ cache integers};

\path (Pi.north west) -- (VG.north east) 
    node[midway, above=1.3cm, font=\Large] {Bit Vector $V$ ($m/\lg^2m$ bits)};

\draw[<->, thick] ([yshift=-0.3cm]Pi.south west) -- ([yshift=-0.3cm]Pi.south east)
    node[midway, below=3pt] {$\lfloor \lg \lfloor 2m/s \rfloor \rfloor + 1$};

\draw[<->, thick] ([yshift=-0.3cm]Vi.south west) -- ([yshift=-0.3cm]Vi.south east)
    node[midway, below=3pt] {$\lfloor \lg g \rfloor + 1$};

\end{tikzpicture}
\caption{Memory layout diagram.}
\end{figure}

\subsection{Structure of
Representatives}\label{structure-of-segment-representatives}

Let \(R\) denote the collection of all \(r_{\#}\)
representatives. We structure \(R\) to support the operations
\(R.\text{search}(x)\) and \(R.\text{insert}(x)\) efficiently. \(R.\text{search}(x)\) finds
the active representative \(R_{i}'\) such that
\(f_{i}' \leq x \leq f_{i + 1}'\), and returns its position in \(R\), or
\(0\) if \(x < f_{1}'\). \(R.\text{insert}(x)\) adds a new active
representative into \(R\) for a newly allocated segment
with \(x\) as its active frame element. Full procedural descriptions and formal analysis of the operations are available in \hyperref[appendix-a-detailed-procedures-and-proofs-for-section-2.2]{[App. A]}.

\(R\) consists of the \emph{library} \(L\) at the front and the
\emph{rebuild buffer} \(D\) at the back.

\textbf{The library} \(L\) contains the sorted sequence of all active
representatives interlaced with buffer representatives, which serve as slack
space for efficient insertions. \(L\) behaves like a dynamic array,
beginning at the left of a reserved zone of
\(8\left\lfloor 2m/s \right\rfloor\) buffer representatives and
expands rightwards.

At any moment, the number of active representatives is \(s_{\#}\) as
there are exactly \(s_{\#}\) active frame elements. Let
\(s_{\#}' =2^{\left\lfloor \lg s_{\#} \right\rfloor}\leq s_\#\) denote the number of active representatives at
the previous resize. \(L\) consists of \(s_{\#}'\) \emph{gaps}
\(L_{1}\ldots L_{s_{\#}'}\), each a block of \(8\)
representatives \(\ell_{1}\ldots \ell_{8}\). Using (2) to bound $|L|$
 \begin{equation*}
|L| \leq 8s_{\#}' \leq 8s_{\#} \leq 8\left\lfloor 2m/s \right\rfloor
\end{equation*}
\noindent
Each \textbf{gap} \(L_{i}\) has the form
\(a_{1}\ldots a_{k}b_{k + 1}\ldots b_{8}\), where \(a_{i}\) are active
representatives in sorted order and \(b_{i}\) are buffer
representatives. The \emph{gap invariant} requires \(0 < k < 8\), so
each gap contains at least one active and one buffer representative. The
sequence \(L_{1}'\ldots L_{s_{\#}'}'\) where \(L_{i}'\) denotes the
active representatives in \(L_{i}\) forms the global sorted sequence of
all active representatives. At startup, \(s_{\#} = 0\), but we
pre-allocate one gap \(L_{1}\) containing only buffer representatives
and set \(s_{\#}' = 1\).

\textbf{The rebuild buffer} \(D\) is a contiguous block of fixed size
\(\left\lfloor 2m/s \right\rfloor_{}^{}\) representatives.
Typically, \(D\) consists only of buffer representatives, but active
representatives are temporarily stored in \(D\) during rebalancing. By
(2), \(|D| > s_{\#}\), so \(D\) can hold every active representative.

This gives the total number of representatives
 \begin{equation}
r_{\#} = |L| + |D| \leq 8\left\lfloor 2m/s \right\rfloor + \left\lfloor 2m/s \right\rfloor = 9\left\lfloor 2m/s \right\rfloor
\end{equation}
\noindent
It follows that the required size of the frame memory, cache memory, and bit vector are respectively
\begin{gather}
|F| = r_{\#} \leq 9\left\lfloor 2m/s \right\rfloor \leq 9m/\lg^{4}m,\\
|G| = r_{\#}2g \approx 18m/\lg^{3}m,\\
|V| = r_{\#}\left( \left\lfloor \lg{\lfloor 2m/s \rfloor} \right\rfloor + \left\lfloor \lg g \right\rfloor + 2 \right) \approx 9m/\lg^{3}m \leq m/\lg^{2}m,
\end{gather}
\noindent
for sufficiently large \(m\). From (3), (6), (7), the required size of
the buffer memory is then
 \begin{equation}
|B| = |F| + |G| + |S| \leq 3m - 1
\end{equation}
\noindent
\textbf{The operation} \(R.\text{search}(x)\) is implemented by two binary searches to locate the final active representative $R_y \leq x$.  The outer search locates the last gap \(L_{i}\) among \(L_{1}\ldots L_{s_{\#}'}\) whose first representative \(\mathcal{\ell}_{1} \leq x\), leveraging the existence of an active representative at the front of every gap by the gap invariant to guarantee a contiguous, monotonic search space. The second binary search then locates the exact last representative $\ell_j \leq x$ within the isolated gap, returning its position. Because $R'_z=R_y$ is the last active representative  $\leq x$, the next active representative $R'_{z+1}>x$. Thus, it obeys the ordering invariant $f'_z \leq x < f'_{z+1}$.

\begin{lemma} \label{L1p1}
\(R.\text{search}(x)\)
requires
\(\lg s_{\#}' + 3 = \left\lfloor \lg s_{\#} \right\rfloor + 3\) comparisons.
\end{lemma}
\begin{proof}
    The outer binary search costs \(\lg s_{\#}'\) comparisons exactly since
    \(s_{\#}'\) is a power of \(2\). The inner search then requires \(\lg 8 = 3\)
    comparisons. Summing gives $\lfloor \lg s_\# \rfloor +3$ comparisons.
\end{proof}

\noindent
\textbf{The operation} \(R.\text{insert}(x)\) is implemented by first locating the position where $x$ belongs by $R.\text{search}(x)$. Then, by a standard array-based linear shift within the gap implemented with exchanges, we shift a buffer representative from the back of the gap $x$ belongs in to the desired location, making room for $x$ to be constructed in its sorted position, maintaining the sorted order of active frame elements. To ensure that a buffer representative is always present at the end of the gap for future insertions, we verify if the gap invariant still holds, executing dynamic rebalancing to restore the invariant if it was violated by the insertion. Occasionally when \(s_{\#} = 2s_{\#}'\) a resize exchanges all active representatives into \(D\) and rebuilds the library structure with \(s_{\#}\) gaps each containing one active representative, updating \(s_{\#}' \coloneqq s_{\#} = 2s_{\#}'\).

\textbf{Dynamic rebalancing} uses an imaginary segment tree over the gaps \(L_{1}\ldots L_{s_{\#}'}\) with nodes \(n_{1}\ldots n_{2s_{\#}' - 1}\),
where \(n_{1}\) is the root node \emph{covering} all gaps, and leaves
\(n_{s_{\#}'}\ldots n_{2s_{\#}' - 1}\) correspond with individual gaps.
The nodes are indexed like a binary tree. For each node \(n_{k}\)
covering the range \(L_{l}\ldots L_{r}\), let
\(\text{size}\left( n_{k} \right)\) denote the number of active representatives
in the range, and let \(\text{span}\left( n_{k} \right) = r - l + 1\), i.e. the
size of the range.
The rebalancing strategy is inspired by lazy weight balancing in
scapegoat trees from \cite{GR93}. When we encounter a
full gap \(L_{i}\), we traverse towards the root from its corresponding
leaf node \(n_{j}\) where \(j = i + s_{\#}' - 1\) (with
\(\text{size}\left( n_{j} \right) = 8\)). At every step of the traversal, a
linear scan counts \(\text{size}\left( n_{j \oplus 1} \right)\) (i.e. the sibling's size), then we
compute 
\(\text{size}\left( n_{\left\lfloor j/2 \right\rfloor} \right) = \text{size}\left( n_{j} \right) + \text{size}\left( n_{j \oplus 1} \right)\),
and set \(j \coloneqq \left\lfloor j/2 \right\rfloor\), moving to the parent.
This process repeats until we encounter a \emph{scapegoat node}
\(n_{j}\) that satisfies\looseness=-1
\begin{equation}
\left| \text{size}\left( n_{2j} \right) - \text{size}\left( n_{2j + 1} \right) \right| > \max\left( 1,\text{size}\left( n_{j} \right)/\lg s_{\#}' \right)
\end{equation}
\noindent
Because the size of the children of a node $n_j$ that does not satisfy (10) are bounded by $\lfloor \text{size}(n_j)/2\rfloor+\max(1,\text{size}(n_j)/\lg s'_\#)$, repeating this linear recurrence down from the root to a leaf bounds the leaf size by $2e+1$ (where $e\approx2.71$ is Euler's constant). But as $L_i$ is full containing $8>2e+1$ active representatives, we know a scapegoat node that satisfies (10) must exist on the path. 

The subtree rooted at the scapegoat node is then rebuilt: move all \(z\) active representatives into \(D\), then
redistribute them evenly across the gaps in the subtree so that for
every internal node \(n_{k}\) in the subtree
\(0 \leq \text{size}\left( n_{2k} \right) - \text{size}\left( n_{2k + 1} \right) \leq 1\).
To compute the block of active representatives for each gap, we conduct
an in-order traversal of the subtree with a stack of one bit per tree
level packed into \(O(1)\) words that allows us to recalculate the
correct block for each ancestor node \hyperref[appendix-a-detailed-procedures-and-proofs-for-section-2.2]{[App.~A]}. Upon reaching a leaf node, we
transport the correct block into the gap. Because the sizes of the scapegoat node's children differ by at least 2 by condition (10), at least one child is forced to fall strictly below its maximum capacity. This limits $\text{size}(n_j) \le 7 \text{span}(n_j)$, ensuring that an even redistribution leaves at most 7 active representatives per gap, restoring the gap invariant. The process rearranges the positions of active representatives, invalidating the location returned by previous $R.\text{search}$ operations.

\begin{lemma} \label{L1p2}
\(R.\text{insert}(x)\) requires
\(O( \lg^{3}m )\) comparisons and
\(O( \lg^{3}m)\) moves amortized.
\end{lemma}

\begin{proof}
Excluding the cost of rebalancing, insertions cost \(O(\lg m)\) amortized: the initial search is $O(\lg m)$, the linear shift conducts $O(1)$ representative exchanges as the size of gaps is constant, and the $2s'_{\#}$ representative exchanges conducted by resizing are spread among the $s'_{\#}/2$ insertions since the previous resize to $O(1)$ representative exchanges per insertion.

To bound rebalancing costs, we use the accounting method. For each internal node \(n_{k}\), let
\(\psi\left( n_{k} \right) = \max\left( 0,\left| \text{size}\left( n_{2k} \right) - \text{size}\left( n_{2k + 1} \right) \right| - 1 \right)\) denote its \emph{imbalance}.
 Now consider an instance of rebalancing that rebuilds a subtree of size $z$. The traversal to find a scapegoat node costs \(z = \text{size}\left( n_{j} \right)\) comparisons, as $1$ comparison is used to count each active representative
\hyperref[appendix-a-detailed-procedures-and-proofs-for-section-2.2]{[App.~A]}. The rebuild then swaps each representative into then out of $z$ using $2z$ representative exchanges costing $O(z \lg m)$, resetting imbalance in the subtree to $0$. By condition (10), prior to the rebuild the subtree contained \(\Omega(z/\lg s_{\#}')\) units of imbalance, so each unit of imbalance is repaired at \(O(z\lg m)/\Omega(z/\lg s_{\#}')=O( \lg^{2}m )\) cost. Inserting a new active representative increases the size of a leaf node by \(1\), which may increase the imbalance of each of its \(\lg s_{\#}'\) ancestors by \(1\). Each unit incurred may later be repaired at \(O(\lg^{2}m)\) cost, bounding the cost of dynamic rebalancing at \(O(\lg^{3}m)\) amortized. Combining costs gives the overall bound.
\end{proof}

\subsection{Sorting Small
Subsequences}\label{sorting-small-subsequences}

We now introduce a specialized \emph{small sort} that is designed to sort small subsequences \(A'\) of \(m' = O(\lg^c m)\) active elements for some constant $c$  given a \emph{merge} \emph{buffer} \(B'\) of
\(m'\) buffer elements, the hole, and \(O(1)\) auxiliary words of
\(w \geq \lg m\) bits, adapted from \cite{KP99}. This subroutine is used during extraction to sort segments. The algorithm is implemented as a \(k\)-way merge sort, where
\begin{equation*}
k = 2^{\left\lceil \lg\left( w/\lg{\lg m} \right) \right\rceil} \geq \lg m/\lg{\lg m}
\end{equation*}
\noindent
It begins by sorting runs of size \(k\) by table sort from
\cite{Knu98} with merge sort on the pointers. Subsequently, adjacent groups of \(k\) runs are merged by using a selection tree to
repeatedly select the minimum element from the current front of all
runs. The choice of $k$ as a power of 2 ensures the selection tree forms a perfect binary tree, minimizing overhead. The pointer array and selection tree occupy
\(O(k)\left( \left\lfloor \lg{m'} \right\rfloor + 1 \right) = O(w)\)
bits, which is packed into \(O(1)\) words. This process repeats until
the entire block \(A'\) is a contiguous sorted run. This requires
\(k_{\#} = \left\lceil \lg{m'}/\lg k \right\rceil - 1\) passes of merging. The use of table sort in the first phase is to avoid the overhead of constructing the selection tree in the second merging phase. For the formal step-by-step procedure and extended analysis, see \hyperref[appendix-b-full-small-sort-procedure]{{[}App. B{]}}.

\begin{lemma} For any constant $c$, a sequence of
\(m' = O(\lg^c m)\) active elements can be
sorted using no more than \(m'\lg{m'} + o(m')\) comparisons and
\(O(m')\) moves if given a merge buffer of \(m'\)
buffer elements.
\label{L2}
\end{lemma}
\begin{proof}
Each merge pass requires at most \(m'\lg k + O( m' )\)
comparisons. In the final pass, there are only
\(k' = \left\lceil m'/k^{k_{\#}} \right\rceil\) runs, so it needs only
\(m'\left\lceil \lg{k'} \right\rceil + O( m' )\) comparisons.
Summing over the initial run sorting and all merge passes, the total
comparison cost satisfies
\begin{align*}
C^{''}\left( m' \right) &= k_\# m' \lg k +m'\lceil\lg \lceil m'/k^{k_\#} \rceil \rceil -m'+o(m') \\
&\leq m' \lg k^{k_\#} +m' \lg (m'/k^{k_\#})+m'-m'+o(m')=m' \lg m' + o(m')
\end{align*}
\noindent
The table sorting uses \(3m'\) moves, and each
merge pass another \(3m'\) moves. Altogether this gives
\(3\left( k_{\#} + 1 \right)m' = O(m')\)
moves, since \(k_{\#} = O(1)\) for $m=O(\lg^c m)$ and
\(w \geq \lg m\).
\end{proof}

\subsection{Inserting into the
Structure}\label{inserting-into-the-structure}

Using the above established components, let us now consider insertion
into the full structure. In this process, we must maintain the ordering invariant ($f'_i\leq x \leq f'_{i+1}$ for all active $x \in (G'_i \cup S_i)$) and the segment invariant (segments contain between $\lfloor s/2 \rfloor$ and $s$ active elements). The structure is initialized accordingly using $O(s)$ moves:
\begin{enumerate}
\def\labelenumi{\arabic{enumi}.}
\item
  Allocate \(S_{0}\) and move the first
  \(\left\lfloor s/2 \right\rfloor\) elements of \(A\) to the front of
  \(S_{0}\) to satisfy the segment invariant. The ordering invariant holds trivially as there is only one segment.
\item
  Create auxiliary pointers \(l_{0}\) to the
  \(\left\lfloor s/2 \right\rfloor + 1\)th position in \(S_{0}\) and
  \(r_{0}\) to the end of \(S_{0}\).
\end{enumerate}

\noindent
Then, for the remaining \(m - \left\lfloor s/2 \right\rfloor\) active elements \(x \in A\) sequentially, we execute the following process.
\begin{enumerate}
\def\labelenumi{\arabic{enumi}.}
\item
  Call \(R.\text{search}(x)\) to locate the proper active representative
  \(R_{y} = R_{z}'\) such that \(f_{z}' \leq x < f_{z + 1}'\). The $S_0$ case is handled separately; here we focus on the general case where \(R.\text{search}(x)\neq0\).
\item
  Insert \(x\) into the cache line of the found representative via
  \(G_{y}.\text{insert}(x)\). The search guarantees
  \(f_{z}' \leq x < f_{z + 1}'\), respecting the ordering invariant for
  \(G_{z}'\) which is \(f_{z}' \leq x \leq f_{z + 1}'\).
\item
  If \(G_{y}.\text{insert}(x)\) returns true, \(G_{y}\) is full and is flushed
  into its corresponding segment \(S_{z}\).
  \begin{enumerate}
  \def\labelenumii{\roman{enumii}.}
  \item
    Load the physical position of \(S_{z}\) into a variable by reading the segment pointer: \(p \coloneqq P_{y}\).
  \item
    By a binary search over the \(s\) positions of \(S_{z}\), find
    the first buffer element \(b_{k}\) in $s$.
  \item
    Call \(G_{y}.\text{flush}(b_{k})\), which moves the \(g\) active
    elements from \(G_{y}\) into the next \(g\) positions after
    \(b_{k}\) in \(S_{z}\).
  \end{enumerate}
This cannot violate the ordering invariant, as it is identical for
\(G_{z}'\) and \(S_{z}\). If the cache line is not full, the procedure
for this insertion terminates immediately. Consequently, from this point onward, we assume the physical position of \(S_{z}\) is kept in the auxiliary variable \(p\), so \(S_{z}\) can be accessed directly.
\item
  If, after the flush, \(S_{z}\) contains \(s\) active elements, it is
  full and must be split into two segments of
  \(\left\lfloor s/2 \right\rfloor\) elements, with the median element
  promoted to the frame. Verifying fullness is done checking if the last element of $S_z$ is active. By construction a segment
  starts with \(\left\lfloor s/2 \right\rfloor\) active elements, so filling a segment requires \(s - \left\lfloor s/2 \right\rfloor = \left\lceil \lg^{3}m \right\rceil g\) additional active elements (see (1)), ensuring whole cache flushes completely fill segments. Thus a flush cannot overflow the segment. The detailed splitting process follows on the next page.\looseness=-1
  
  \begin{figure}
\centering
\begin{tikzpicture}[
    >=Straight Barb,
    cell/.style={draw, thick, minimum height=1cm, inner sep=0pt},
    font=\Large,
    scale=0.7, transform shape 
]

\colorlet{movecolor}{gray!20}

\node[cell, minimum width=3cm] (le) {$\le a_\text{med}$};
\node[cell, minimum width=1.5cm, right=0pt of le] (am) {$a_\text{med}$};
\node[cell, minimum width=3cm, right=0pt of am, fill=movecolor] (ge) {$\ge a_\text{med}$};
\node[below=0.4cm of am, font=\huge] {$S_z$};

\node[cell, minimum width=3cm, right=3cm of ge, fill=movecolor, dashed] (sz1front) {$\ge a_\text{med}$};
\node[cell, minimum width=4.5cm, right=0pt of sz1front, draw=none] (sz1back) {};

\draw[thick] (sz1front.north west) rectangle (sz1back.south east);

\path (sz1front.north west) -- (sz1back.north east) node[midway, above=0.4cm, font=\huge] {$S_{z+1}$};

\node[above right=1.5cm and 1cm of ge.north east, font=\huge] (f) {$f'_{z+1}$};

\draw[thick] (f.south) -- ++(0, -3);

\draw[->, thick, shorten >=2pt] (am.north) to[out=60, in=180] (f.west);
\draw[->, thick, shorten >=2pt] (ge.south) to[out=-60, in=-120] (sz1front.south);

\end{tikzpicture}
\caption{Movement of active elements during splitting a full segment $S_z$.}
\end{figure}
  \begin{enumerate}
  \def\labelenumii{\roman{enumii}.}
  \item
    Select the median element $a_\text{med}$ of \(S_{z}\) by the linear-time stable, in-place selection algorithm of \cite{KP92a} then stably ternary partition $S_z$ around pivot $a_\text{med}$ \cite{KP92b}, logically dividing the block $=a_\text{med}$ at the midpoint of $S_z$. This leaves the elements of $S_z$ in the configuration
    \(a_{1}\ldots a_{\left\lfloor s/2 \right\rfloor}a_\text{med}a_{1}'\ldots a_{\left\lfloor s/2 \right\rfloor}'\)
    where \(a_{i} \leq a_\text{med} \leq a_{i}'\) for all
    \(1 \leq i \leq \left\lfloor s/2 \right\rfloor\).
  \item
    Allocate the new segment \(S_{z + 1}\), which returns a pointer to its start.
  \item
    Swap the block \(a_{1}'\ldots a_{\left\lfloor s/2 \right\rfloor}'\)
    to the front of \(S_{z + 1}\).
  \item
    Create the active representative for \(S_{z + 1}\) by
    \(R.\text{insert}(a_{med})\). This may invalidate the location returned by $R.\text{search}$  in step 1 (see \hyperref[structure-of-segment-representatives]{Section 2.2}). However, as the insertion terminates immediately after without accessing representatives, correctness is unaffected.\looseness=-1
  \end{enumerate}
  \vspace{-1em}
  After splitting, the ordering invariants are maintained by construction. Since all original elements \(x \in S_{z}\) satisfied \(f_{z}' \leq x \leq f_{z + 1}'\),
the elements $\leq a_{med}$ remaining in \(S_{z}\) satisfy
\(f_{z}' \leq x \leq a_{med}\) while those moved to \(S_{z + 1}\) are from the other block $\geq a_{med}$ and satisfy \(a_{med} \leq x \leq f_{z + 1}'\). After \(a_{med}\) is promoted as the new \(f_{z + 1}'\), the invariants
\(f_{z}' \leq a_{i} \leq f_{z + 1}'\) and
\(f_{z + 1}' \leq a_{i}' \leq f_{z + 2}'\) hold true. Similarly, the segment invariant is respected: $S_{z+1}$ receives $\lfloor s/2 \rfloor$ elements and $s-\lfloor s/2 \rfloor -1=\lfloor s/2 \rfloor$ elements (see (1)) remain in $S_z$.
\end{enumerate}

\noindent
In the case where the $R.\text{search}(x)=0$ in step 1 directing us to \(S_0\), we conduct the following:
\begin{enumerate}
\def\labelenumi{\arabic{enumi}.}
\item
  Swap \(x\) with the element at position \(l_{0}\), then increment
  \(l_{0}\).
\item
  If \(l_{0} > r_{0}\), \(S_{0}\) is full and split using the procedure above, then update \(l_{0}: = l_{0} - \left\lceil s/2 \right\rceil\)
  so $l_0$ points to the next available buffer space in \(S_{0}\) after the
  split.
\end{enumerate}
\vspace{-0.25em}
\enlargethispage{0.02em}
\begin{lemma}
\label{L3}
Inserting all \(m\) active elements into the structure
requires no more than \(m\lg\left\lfloor m/s \right\rfloor{} + O(m)\)
comparisons, \(O(m)\) moves, \(m\) cache inserts
and \(\left\lfloor m/g \right\rfloor\) cache flushes.
\end{lemma}

\begin{proof}
We now analyze the cost of inserting all \(m\) elements into the structure step-by-step.
\begin{enumerate}
\def\labelenumi{\arabic{enumi}.}
\item
  \(R.\text{search}(x)\) requires
  \(\left\lfloor \lg s_{\#} \right\rfloor + 3 \leq \lg\left\lfloor m/s \right\rfloor + 4\)
  comparisons by \hyperref[L1p1]{Lemma 2.1} and (2).
\item
  Inserting \(x\) into the cache line requires a single cache insert. Over \(m - \left\lfloor s/2 \right\rfloor \leq m{}\) insertions the first 2 steps
cost \(m\lg\left\lfloor m/s \right\rfloor + 4m\) comparisons and \(m\)
cache inserts.
\item
  Cache lines are flushed only after \(g\) insertions into the same
  cache line, so cache flushes occur at most
  \(\left\lfloor m/g \right\rfloor\) times, each costing \(O(\lg m)\)
  comparisons and a single cache flush, totaling \(O(m)\) comparisons
  and \(\left\lfloor m/g \right\rfloor\) cache flushes.
\item
  Splitting is triggered when a segment has become full after the cache flush.
  Every split requires: $1$ comparison to check fullness, $O(1)$ arithmetic operations to allocate a new segment, \(O(s)\) comparisons and \(O(s)\) moves for
  stable selection and partitioning by the median \cite{KP92a,KP92b},
  \(3\lfloor s/2\rfloor\) moves to transport
  \(a_{1}'\ldots a_{\left\lfloor s/2 \right\rfloor}'\) to the new
  segment, and \(O( \lg^{3}m ) = o(s)\) comparisons and
  \(O( \lg^{3}m ) = o(s)\) moves for \(R.\text{insert}(a_{med})\) by
  \hyperref[L1p2]{Lemma 2.2} and (1). In total, a split costs \(O(s)\) comparisons and
  \(O(s)\) moves. Since each split creates a new segment, and there are
  \(s_{\#}\ \) additional segments, there have been \(s_{\#}\) splits, costing
  \(s_{\#}O(s) \leq O(m)\ \)comparisons and \(s_{\#}O(s) \leq O(m)\) moves.
\end{enumerate}

\noindent
Inserting an element into \(S_{0}\) is equivalent to a standard insertion without the overhead of cache operations (steps 2, 3) by leveraging the auxiliary pointers. Thus, the cost lies within the general bound. Adding the per-step bounds yields the bounds stated in the lemma.
\end{proof}

\subsection{Extracting in Sorted Order}\label{extracting-in-sorted-order}

\(\)The extraction phase transports all active elements back to \(A\) in
sorted order. After insertion completes, every active element lies either in the
frame, cache or segments, whilst \(A\) contains only buffer elements.

By a linear scan through the \(L\), for every active representative $R'_i$,
\(G_{i}'\) is emptied into \(S_{i}\) by the same method as step 3 of insertion. In total,
\(o(m)\) comparisons and \(s_{\#}\) cache flushes are required to empty
all cache lines. This leaves active elements either as active frame
elements \(f_{1}'\ldots f_{s_{\#}}'\) in sorted order or in segments
\(S_{0}\ldots S_{s_{\#}}\), where for any \(a_{i} \in S_{i}\), the
ordering invariant \(f_{i}' \leq a_{i} \leq f_{i + 1}'\) holds. It
follows that, the sequence
\(S_{0}'f_{1}'S_{1}'f_{2}'\ldots f_{s_{\#}}'S_{s_{\#}}'\) is the sorted
sequence of all \(m\) active elements, where \(S_{i}'\) denotes the
sorted sequence of all active elements in \(S_{i}\). We now construct
this sequence in \(A\).

First, we transport \(S_{0}'\) to the front of \(A\). We sort the active
elements in \(S_{0}\) (from \(l_{0}\) to \(r_{0}\)) with the small sort
using \(A\) as the merge buffer, then move the sorted block to the front
of \(A\).

Next, we linearly scan through \(L\) from left to right. for each active
representative \(R_{i}'\) we encounter, we conduct the following:
\begin{enumerate}
\def\labelenumi{\arabic{enumi}.}
\item
  Swap the element \(f_{i}'\) with the next available buffer space in
  \(A\).
\item
  Read the corresponding active segment pointer \({p \coloneqq P}_{i}'\) to
  locate the physical position of \(S_{i}\). \looseness=-1
\item
  Perform a binary search over \(S_{i}\) to find the number of active
  elements in \(S_{i}\), denoted by \(s_{i}\). Let the active elements in $S_i$ be \(a_{1}\ldots a_{s_{i}}\).
\item
  Using \(S_{0}\) (already emptied) as the merge buffer, sort
  \(a_{1}\ldots a_{s_{i}}\) with the small sort.
\item
  Move \(a_{1}\ldots a_{s_{i}}\) into the next \(s_{i}\) buffer
  positions in \(A\).
\end{enumerate}

\noindent
Every iteration adds the sequence of active elements \(f_{i}'S_{i}'\) to
the end of the growing sorted sequence in \(A\). Since active
representatives appear in sorted order in \(L\), the resulting sequence
in \(A\) is exactly
\(S_{0}'f_{1}'S_{1}'f_{2}'\ldots f_{s_{\#}}'S_{s_{\#}}'\), the global
sorted sequence of all \(m\) active elements.

Finally, to fulfill the Section 2 termination requirements, we verify the ending states of $B$ and $V$. Throughout, the algorithm writes elements only by swapping (via \(\text{hole} \coloneqq a\), \(a \coloneqq b\), \(b \coloneqq \text{hole}\)), which exclusively permutes elements. Since the final sorted permutation of all active elements is in \(A\), then the remaining elements in \(B\) are already a permutation of the original buffer elements. However, the bit vector remains modified, requiring a linear scan to reset it to all $\texttt{false}$, using \(|V| = o(m)\) comparisons and at most \(3|V| = o(m)\) moves (see (8)). Subsequently the algorithm terminates safely.

\begin{lemma}
\label{L4}\ Extracting
all \(m\) active elements from the structure
in sorted order requires no more than \(m\lg s + O(m)\)
comparisons, \(O(m)\) moves and \(s_{\#}\) cache
flushes.
\end{lemma}

\begin{proof}
The linear scan in \(L\) requires \(|L|=8s_{\#}'\) comparisons. Each segment is processed using $s_i \lg s_i+O(s_i)$ comparisons and $O(s_i)$ moves (handling $S_0$ is a simplified case that lies within the same bounds):
\begin{enumerate}
\def\labelenumi{\arabic{enumi}.}
\item
  Swapping \(f_{i}'\) into \(A\) costs \(3\) moves.
\item
  reading \(P_{i}'\) requires
  \(\left\lfloor \lg{m/s} \right\rfloor + 2\) comparisons.
\item
  The binary search for \(s_{i}\) takes
  \(\left\lfloor \lg s \right\rfloor + 1\) comparisons.
\item
  Sorting the segment requires \(s_{i}\lg s_{i} + o(s_{i})\) comparisons
  and \(O(s_{i})\) moves by \hyperref[L2]{Lemma 2.3}.
\item
  Swapping \(s_{i}\) elements into \(A\) costs \(3s_{i}\) moves.
\end{enumerate}

\noindent
Let $C_\text{ext}(m)$ and $M_\text{ext}(m)$ denote the comparisons and moves respectively made during extraction. As $s_i \leq s$ and \(\sum_{i = 0}^{s_{\#}}s_{i} \leq m\), summing over all segments gives
\begin{equation*}
\begin{array}{l}
  C_\text{ext}(m) = 8s_{\#}' + \sum_{i = 0}^{s_\#}(s_{i}\lg s_{i} + O( s_{i} )) \leq 8s'_\# + \sum_{i = 0}^{s_\#}(s_{i} ( \lg s + O(1) ) \leq m \lg s + O(m)
  \\[1ex]
  M_\text{ext}(m)=\sum_{i = 0}^{s_{\#}}{O( s_{i} )} = O(m)
\end{array}
\qedhere
\end{equation*}
\end{proof}

\subsection{Structure of Cache
Lines}\label{structure-of-cache-lines}

We now specify the two different designs for cache lines (defined in
\hyperref[structure-of-the-memory]{Section 2.1}).

First, consider the \emph{randomized cache line} used in the \textbf{randomized approach}. We choose
\(g = 2^{\left\lceil \lg{\lg m} \right\rceil}\), which satisfies (4).
Let us denote: the cache line as the block of elements
\(g_{1}\ldots g_{2g}\), and the corresponding cache integer as \(d\), with
bits \(d_{0}\ldots d_{\lg g}\), where bit
\(d_{i}\) has weight \(2^{i}\). This is well defined as $g$ is a power of 2.

The process of \(\text{insert}(x)\) in a randomized cache line works as
follows:
\begin{enumerate}
\def\labelenumi{\arabic{enumi}.}
\item
  Probe for a buffer slot. Repeatedly select a uniformly random index
  \(j \in \lbrack 1\ldots 2g\rbrack\), check whether \(g_{j}\) is a
  buffer element, and retry if it is active until a buffer position
  \(g_{j}\) is found.
\item
  Swap the active element \(x\) with \(g_{j}\), placing \(x\) in the
  cache line.
\item
  Increment \(d\) to reflect the new active element.
\item
  Return the state of the cache line (full or not), given by the value
  of the bit \(d_{\lg g}\) with priority \(g\).
\end{enumerate}

\noindent
Randomized cache lines store data sparsely like hash tables for constant-time insertions: at most \(g\) active elements are spread among \(2g\) slots, capping the load factor at \(1/2\), so the
expected number of probes is \(1/(1 - 1/2) = 2\), each requiring one
comparison. Then, the swap requires \(3\) moves. The increment of \(d\)
is \(O(1)\) amortized by aggregate analysis. Finally, checking for
fullness requires reading the bit \(d_{\lg g}\) using
one comparison. In total, each insertion requires \(O(1)\) comparisons expected and
\(O(1)\) moves amortized (see \hyperref[appendix-c-analysis-of-cache-lines]{[App. C]} for a more comprehensive analysis).

The operation \(G_{i}.\text{flush}(x)\) is a simple linear scan with swaps.
Iterating through \(g_{1}\ldots g_{2g}\), identify every active element
\(g_{j}\), and swap \(g_{j}\) with the next available buffer space after
\(x\). Lastly, reset the cache integer by \(d \coloneqq 0\). This process
uses \(2g\) comparisons and \(\leq3g + 3\) moves. \looseness=-1

\begin{theorem}
\label{T6p1} A
sequence \(A\) of \(m\) active elements can be sorted
using \(m\lg m + O(m)\) comparisons with exponentially high probability and
always \(O(m)\) moves if given a buffer memory \(B\) containing $3m-1$ buffer elements
and bit vector \(V\) of size $m/\lg^2 m$.
\end{theorem}
\begin{proof}
By instating the randomized cache line design, cache insertions cost $O(1)$ comparisons expected and $O(1)$ moves amortized and cache flushes cost $O(g)$ comparisons and $O(g)$ moves (shown directly above). Substituting these costs into Lemmas 2.4 and 2.5 then summing them gives $m\lg m + O(m)$ comparisons expected and $O(m)$ moves deterministically. Over \(m\) cache insertions, the probability of large deviations (i.e. $>m$ comparisons above the expected amount) is bounded by $(27/32)^m$ (see the Chernoff tail bound in \hyperref[appendix-c-analysis-of-cache-lines]{{[}App.
C{]}}).
\end{proof}

\noindent
Next, we detail the mechanics of \emph{deterministic cache lines} for the \textbf{deterministic approach}.
Deterministic cache lines consist of \(t'\) \emph{cache blocks}
\(G_{t' + 1}\ldots G_{1}\) from left to right, where \(g_{i} = |G_{i}|\)
(we locally redefine \(G_{i}\) and \(g_{i}\) here). The block \(G_{i}\)
consists of \(g_{i}/g_{i - 1}\) \emph{sub cache blocks}
\(A_{1}\ldots A_{j}B_{j + 1}\ldots B_{g_{i}/g_{i - 1}}\) each of size
\(g_{i - 1}\), where \(A_{1}\ldots A_{j}\) contain only active elements
and \(B_{j + 1}\ldots B_{g_{i}/g_{i - 1}}\) contain only buffer
elements.

The sizes are controlled by a parameter \(t\) that satisfies
\(2 \leq \left\lceil \lg^{(t)}m \right\rceil \leq \lg{\lg m} + 1\). They
are defined as \(g_{0} = 1\),
\(g_{1} = 2^{\left\lceil \lg^{(t)}m \right\rceil}\), and
\(g_{i} = g_{i - 1}2^{g_{i - 1}/2^{i - 1}}\) for all
\(2 \leq i \leq t'\), where \(t'\) is the largest integer such that
\(g_{t'} < \lg m\). To account for any remaining space, we define the
final term (overrides $g_1$ base case when $t'=0$):
\begin{equation*}
g = g_{t' + 1} = g_{t'}\left\lceil \lg m/g_{t'} \right\rceil \leq g_{t'}2^{g_{t'}/2^{t' - 1}}
\end{equation*}

\noindent
This ensures \(\lg m \leq g \leq 2\lg m\) as in (4), and that cache lines take up
no more than \(2g\) spaces
\hyperref[appendix-c-analysis-of-cache-lines]{{[}App. C{]}}. The
recurrence makes \(g_{i}\) grow as a tower of exponentials, so the
number of levels is proportional to \(t\). It can be shown \(t' \leq t - 1\) by
induction \hyperref[appendix-c-analysis-of-cache-lines]{[App. C]}. This growth also guarantees that the amortized
comparison cost at levels \(i \geq 2\) form a convergent geometric
series, shown shortly.

\(\)The procedure for \(\text{insert}(x)\) iterates through levels
\(i = 1\ldots t' + 1\). At every level:
\begin{enumerate}
\def\labelenumi{\arabic{enumi}.}
\item
  Binary search in \(G_{i}\) for the next available sub cache block
  \(B_{j + 1}\) of buffer elements.
\item
  Swap the entire contents of the lower leveled cache block
  \(G_{i - 1}\) (or simply \(x\) when \(i = 1\)) into the found buffer
  block.
\item
  If this fills \(G_{i}\) (i.e. the last buffer block
  \(B_{g_{i}/g_{i - 1}}\) was used), we continue to level \(i + 1\),
  otherwise insertion completes returning 0.
\end{enumerate}

\noindent
If the loop ends without \(\text{insert}\) terminating, the last block
\(G_{t' + 1}\) has been filled, and the cache line is full thus the
procedure returns \(1\).

At level \(i\), the \(\lg{(g_{i}/g_{i - 1})}\) comparisons to locate a
buffer block by binary search is spread among \(g_{i - 1}\) elements to
\(\lg\left( g_{i}/g_{i - 1} \right)/g_{i - 1}\) comparisons per element.
For \(i \geq 2\), the recurrence yields
\(\lg\left( g_{i}/g_{i - 1} \right)/g_{i - 1} \leq 1/2^{i - 1}\).
Summing over all levels, the total comparisons per element are
\(\lg g_{1} + \sum_{i = 2}^{t' + 1}{1/2^{i - 1}} < \lg^{(t)}m + 2\).
Each element passes through every level once by a single swap costing
\(3\) moves, which over all levels is \(3\left( t' + 1 \right) \leq 3t\)
moves.

The operation \(G_{i}.\text{flush}(x)\) iterates through all cache blocks
\(G_{t' + 1}\ldots G_{1}\) in descending order of level, where for each
cache block a binary search locates last sub cache block
containing active elements \(A_{j}\), swapping all active elements in
the cache block in \(A_{1}\ldots A_{j}\) to the next available buffer
spaces after \(x\). Because each level has \(O(\lg m)\) sub cache
blocks, each binary search costs \(O(\lg{\lg m})\) comparisons. With
\(t' = O(\lg^{*}m)\) levels, the total comparisons are
\(O(\lg{\lg m} \cdot \lg^{*}m)\). Each active element is swapped exactly once, totaling \(3g\) moves.

\begin{theorem}
\label{T6p2}{} A
sequence \(A\) of \(m\) active elements can be sorted
using no more than \(m\lg m + O(m\lg^{(t)}m)\) comparisons
and \(O(tm)\) moves if given a buffer memory \(B\) containing $3m-1$ buffer elements and bit vector \(V\) of size $m/\lg^2 m$.
\end{theorem}
\begin{proof}
    The result follows by a parallel derivation to Theorem~\ref{T6p1}, substituting the deterministic cache operation costs (shown just above) into the bounds of Lemmas 2.4 and 2.5.\looseness=-1
\end{proof}

\begin{remark}
The deterministic additional memory algorithm is stable on the active elements, i.e. the relative order of equal active elements in the output is identical to the input. This does not apply to the randomized variant as active elements are placed arbitrarily in randomized cache lines. However, any in-place adaptation is unstable as the buffer memory is simulated by elements in the array, which are scrambled arbitrarily. If the buffer memory does not contain any equal array elements (e.g. if it is external or an explicitly collected distinct buffer), global stability holds. A full proof of this claim is in \hyperref[appendix-c-analysis-of-cache-lines]{[App.
C]}.
\end{remark}

\subsection{Handling Short Blocks}\label{handling-short-blocks}

During the analysis, we have assumed  \(m\) is sufficiently large. When \(m < 2^{19} = 524288\), we use table sort from \cite{Knu98} applying merge sort on pointers
 as a fallback. This uses \(O(1)\) words as \(m\) is constant, albeit
large, and works within the bounds of Theorems 2.6 and 2.7.

\section{Sorting In-Place}\label{sorting-in-place}

Finally, we demonstrate how the buffer memory and bit vector used in \hyperref[sorting-with-additional-memory]{Section 2} are carved out directly from the input array, following the in-place conversion technique of \cite{FG05}.

\textbf{The bit vector} is first created by using heapselect to isolate the smallest and largest $v=\lceil n/\lg^2 n \rceil$ elements into the blocks $L=l_1\dots l_v$ and $R=r_1\dots r_v$ respectively, resulting in the configuration $L\mathcal{A}'R$. Because $L$ and $R$ contain the smallest and largest elements in $\mathcal{A}$, sorting $\mathcal{A}'$ would sort $\mathcal{A}$. This process requires $O(n+v \lg n)=O(n)$ time. The $i$th bit is encoded by the pair $(l_i,r_i)$: if $l_i<r_i$ the bit is 0, otherwise it is 1. The operation $\text{write}(i)$ simply swaps $l_i$ and $r_i$. For this encoding to be valid, no pair can be equal, thus we test $l_v<r_1$: if this is true the encoding is valid, otherwise the region $\mathcal{A}'$ contains only one unique key and $\mathcal{A}$ is already sorted, allowing the algorithm to terminate immediately. As $n\geq m$, $\lceil n/\lg^2 n \rceil \geq m/\lg^2 m$, thus the bit vector is large enough to sort any subsequence of $\mathcal{A}$ using the algorithm from \hyperref[sorting-with-additional-memory]{Section 2}.

\textbf{The buffer memory} is then repeatedly created by a partition based loop. At the $i$th iteration, let $\mathcal{A}'$ be subdivided into two regions $A_SA_U$, where all elements in $A_S$ are in their correct sorted positions, and sorting $A_U$ would sort $\mathcal{A'}$. Initially $A_S$ is empty. At the $i$th iteration, let the size of $A_U$ be $n_i$. Using the linear time in-place selection algorithm from \cite{GK06}, we select a pivot $p$ with rank $\lceil n_i/4 \rceil$ and partition $A_U$ by $p$, resulting in the regions $A_<pB_\geq$ where $A_<$ contains $n_{i,<}$ elements $<p$ and $B_\geq$ contains elements $\geq p$. Then, using $B_\geq$ as the buffer memory and simulating $V$ with the blocks $L$ and $R$, we sort $A_<$ by the additional memory algorithm from \hyperref[sorting-with-additional-memory]{Section 2}. Due to the classification of active and buffer elements relative to $p$, distinguishing active from buffer elements can be done by comparing against $p$: $\text{active}(x)$ simply needs to return the Boolean result of $x<p$. Because $|A_<|\leq\lceil n_i/4\rceil-1$ and $|B_\geq|\geq n_i-\lceil n_i/4\rceil$, $|B_\geq|\geq3|A_<|-1$, the buffer size is sufficient. Lastly, we isolate all elements $=p$ in $B_\geq$, resulting in the two blocks $P$ containing elements $=p$ and $B_>$ holding the remaining elements. In the resulting sequence, the blocks $A_<P$ are already in their correct sorted positions, thus we begin the next iteration with $B_>=A_U$. In each iteration, the selection and partitioning steps require $O(n_i)$ comparisons and $O(n_i)$ moves. Eventually, when $|A_U| < 2^{19} = 524288$, the entire block $A_U$ is sorted by table sort from \hyperref[handling-short-blocks]{Section 2.7} as a base case, and the algorithm terminates as $\mathcal{A}$ is sorted.

\begin{theorem}
The input array 
\(\mathcal{A}\), consisting of \(n\) elements,
can be sorted in-place using w.e.h.p.
\(n\lg n + O(n)\) comparisons, always \(O(n)\) moves, and
\(O(n\lg n)\) arithmetic operations.
\end{theorem}
\begin{proof}
The choice of the pivot $p$ guarantees $n_{i+1}\leq3/4 n_i$, meaning the size of $A_U$ decays geometrically: $\sum_{i=1}^{I} n_i = O(n)$, where $I$ is the number of iterations. Furthermore, because $A_U$ shrinks by more than $n_{i,<}$ elements every iteration, $\sum_{i=1}^{I} n_{i,<} \leq n$. Letting $C'(m)$ and $M'(m)$ denote the number of comparisons and moves respectively for the additional memory algorithm to sort $m$ elements, we sum for the comparisons $C(n)$ and moves $M(n)$ to sort $\mathcal{A}$:
\begin{equation}
\begin{array}{l}
    C(n) = O(n)+\left( \sum_{i = 1}^{I}{C'\left( n_{i, <} \right) + O( n_{i} )} \right)\leq C'(n)+O(n), \\[1ex]
    M(n) = O(n)+\left( \sum_{i = 1}^{I}{M'\left( n_{i, <} \right) + O( n_{i} )} \right)\leq M'(n)+O(n)
\end{array}
\end{equation}

\noindent
The detailed calculation is in \hyperref[appendix-d-in-place-conversion]{[App. D]}. Substituting $m\lg m+O(m)$ comparisons and $O(m)$ moves from Theorem \hyperref[T6p1]{2.6} yields the stated costs. Furthermore, as all arithmetic operations correspond with comparisons and moves, they are bounded by $O(C(n)+M(n))=O(n \lg n)$.

The w.e.h.p. property is also inherited: the same Chernoff bound in \hyperref[appendix-c-analysis-of-cache-lines]{[App. C]} holds globally for cumulative failed probes across all iterations. The probability of large deviations ($>n$ more comparisons than the expected amount) remain bounded by $(27/32)^n$ \hyperref[appendix-d-in-place-conversion]{[App. D]}.\looseness=-1
\end{proof}

\begin{theorem}
The input array
\(\mathcal{A}\), consisting of \(n\) elements, can
be sorted in-place using no more than \(n\lg n + O(n\lg^{(t)}n){}\) comparisons, \(O(tn)\) moves, and \(O(n\lg n)\)
arithmetic operations.
\end{theorem}
\begin{proof}
When performing the in-place conversion across varying block sizes $m$, small instances may violate the parameter constraint $2 \leq \lceil \lg^{(t)} m \rceil \leq \lg\lg m + 1$. To ensure the deterministic scheme remains valid for every subproblem, we set $g_1 = 2^{\max(2, \lceil \lg^{(t)} m \rceil)}$ for each instance sorting $m$ active elements. This adjustment adds at most a linear number of comparisons per subproblem, preserving the asymptotic performance of Theorem~\ref{T6p2}.

Mirroring Theorem~3.1, substituting $C'(m) = m\lg m + O(m\lg^{(t)}m)$ and $M'(m) = O(tm)$ from Theorem~2.7 into (11) yields $n\lg n + O(n\lg^{(t)}n)$ comparisons and $O(tn)$ moves. Similarly, arithmetic operations are also bounded by $O(n \lg n)$. These upper bounds hold unconditionally as Theorem 2.7 provides worst-case bounds.
\end{proof}

\section{Concluding
Remarks}\label{concluding-remarks}
In this paper, we have addressed the exact operational complexity of in-place sorting, establishing that the fundamental lower bound of $n\lg n + \Omega(n)$ total operations is tight for in-place randomized algorithms (with exponentially high probability). By resolving the previous inefficiencies present in \cite{GG11} via introducing a novel ordered set, optimized cache structures, and an optimal small sort, we have achieved the ultimate goal stated in \cite{GG11,MR92} with exponentially high probability: sorting in-place using only \(n\lg n + O(n)\) comparisons and \(O(n)\) moves.

We leave as an open problem the task of retaining the bound of \(n\lg n + O(n)\) total operations in the worst case. The deterministic cache line stands as the only remaining obstacle to optimality, limiting the deterministic approach at \(n\lg n + O(n\lg^{(t)}n)\) comparisons and $O(tn)$ moves (where $t$ is an integer parameter satisfying $2\leq t \leq \lg^* n-1$), which results in a slightly looser total operation bound of $n \lg n + O(n \lg^* n)$ when $t \approx \lg^* n$. However, this is sufficient to answer the open problem from \cite{GG11} for improvements on their upper bound of \(n\lg n + O(n\lg{\lg n})\) comparisons and $O(n)$ moves.

All in-place algorithms are unstable as the buffer elements are permuted arbitrarily. Whilst the deterministic additional memory algorithm is stable on active elements, it requires a completely distinct buffer to operate stably and in-place. As existing methods of collecting such a distinct buffer incur $\Omega(n \lg n)$ comparisons of overhead \cite{Fra07}, we have not yet achieved a stable, in-place sorting algorithm that makes \(n\lg n + o(n\lg n)\) comparisons and \(O(n)\) moves (i.e. $n \lg n + o(n \lg n)$ total operations). That problem remains open. The best known stable bound stands at \(O(n\lg n )\) comparisons and $O(n)$ moves \cite{Fra07}.

\section*{AI Disclosure}
AI tools were used exclusively for: (i) standard copy-editing tasks, including language refinement of the authors' own text (Gemini, Perplexity), and (ii) to assist in converting the authors' hand-drawn diagrams into TikZ code (Gemini). No additional diagrammatic details were introduced by the AI. The technical aspects, including all algorithms and proofs, were formulated entirely by the authors, who assume responsibility for the entire content.


\clearpage
\bibliography{references}

\clearpage
\appendix
\numberwithin{equation}{section}

\section{Detailed Procedures and
Proofs for Representatives Structure}\label{appendix-a-detailed-procedures-and-proofs-for-section-2.2}

Recall that there are \(r_{\#}\) segment
representatives out of which \(s_{\#}\) are active, distributed among
\(s_{\#}'\) gaps \(L_{1}\ldots L_{s_{\#}'}\), each a contiguous block of
\(8\) representatives. Additionally, we also have the rebuild
buffer \(D\) that is large enough to hold all active representatives.
Our objective is to support \(R.\text{search}(x)\) and
\(R.\text{insert}(x)\) efficiently.

\subsection{Search}
The operation \(R.\text{search}(x)\) finds the active representative \(R_{z}'\)
such that \(f_{z}' \leq x < f_{z + 1}'\), and is implemented as two
binary searches:
\begin{enumerate}
\def\labelenumi{\arabic{enumi}.}
\item
  By an outer binary search in \(L_{1}\ldots L_{s_{\#}'}\), locate the
  last gap \(L_{i}\) whose first active representative
  \(\mathcal{\ell}_{1}\) satisfies \(\mathcal{\ell}_{1} \leq x\).
\item
  By an inner binary search over
  \(L_{i} = \ell_{1}\ldots \ell_{8}\), find the first
  representative \(l_{j}\) that is either an active
  representative \(> x\) or a buffer representative (if no such active
  representative exists). The gap invariant guarantees such a segment
  representative exists (as a buffer representative must exist in the
  gap).
\item
  If \(i = 1\) and \(j = 1\), then \(x < f_{1}\), thus \(x\) belongs in
  \(S_{0}\); return 0. Otherwise, the correct active representative is
  \(\ell_{j - 1}\), or \(R_y\) where \(y = 8(i - 1) + (j - 1)\) in
  the global sequence \(R\); return \(y\).
\end{enumerate}

\noindent
Let the resulting active representative from \(R.\text{search}(x)\) be
\(R_{z}'\). We show that \(f_{z}' \leq x < f_{z + 1}'\). By the outer
binary search, \(x\) lies between the first active representatives of
\(L_{i}\) and \(L_{i + 1}\). By the inner binary search, \(\ell_{j}\) is
either the first active representative \(> x\) or the first buffer
representative in \(L_{i}\). Therefore, \(\ell_{j - 1}\) is the last active
representative \(\leq x\). The next active representative is either
\(\ell_{j}\) which is \(> x\) by the inner binary search or the first
representative of \(L_{i + 1}\) which is also \(> x\) by the
outer binary search. Thus, the inequality \(f_{z}' \leq x < f_{z + 1}'\)
holds. \looseness=-1

The outer binary search requires \(\lg s_{\#}'\) comparisons exactly as
\(s_{\#}'\) is a power of \(2\). The inner binary search requires
exactly \(\lg 8 = 3\) iterations as \(8\) is a power of 2. Naively, for
each iteration, \(2\) comparisons are required to first determine if
\(\mathcal{\ell}_{j}\) is a buffer representative, then optionally an
additional comparison against \(x\) for active representatives. However,
as buffer elements are greater than all active elements (see
\hyperref[sorting-in-place]{Section 3}), a single comparison
\(\mathcal{\ell}_{j} > x\) suffices. Lastly, the edge case check and
position calculation requires \(O(1)\) arithmetic operations.

\begin{lemma}
\label{LAp1}
\(R.\text{search}(x)\) requires
\(\lg s_{\#}' + 3 = \left\lfloor \lg s_{\#} \right\rfloor + 3\)
comparisons.
\end{lemma}

\subsection{Insertion}
The procedure for \(R.\text{insert}(x)\) is as follows:
\begin{enumerate}
\def\labelenumi{\arabic{enumi}.}
\item
  Run \(y = R.\text{search}(x)\) to find the representative
  \(\mathcal{\ell}_{j}\) from \(L_{i}\)
  (\(i = \left\lceil y/8 \right\rceil\), \(j = y - 8(i - 1)\)), then
  increment \(j\) giving position where \(x\) should be inserted.
\item
  By a binary search in \(L_{i}\) find the last active representative
  \(a_{k}\) to determine the value of \(k\), i.e. the number of active
  representatives in \(L_{i}\).
\item
  Iterating through \(j' = k\ldots j\) (descending order), we swap every
  \(\mathcal{\ell}_{j'}\) with \(\mathcal{\ell}_{j' + 1}\). This shifts the
  active representatives initially at
  \(\mathcal{\ell}_{j}\ldots\mathcal{\ell}_{k}\) one unit rightwards into the
  positions \(\mathcal{\ell}_{j + 1}\ldots\mathcal{\ell}_{k + 1}\), and
  \(\mathcal{\ell}_{j}\) now contains the representative initially
  at \(\mathcal{\ell}_{k + 1}\), which is a buffer representative (as
  \(\mathcal{\ell}_{k}\) is defined as the last active representative).
\item
  Construct the new active representative in \(\mathcal{\ell}_{j}\),
  promoting \(x\) to the frame.
\item
  Check if the gap invariant was violated, by verifying if \(k = 7\), as
  the insertion would add an additional active representative, creating
  a gap with \(8\) active representatives, violating the invariant. If
  so, execute the dynamic rebalancing procedure on \(L_{i}\), which
  redistributes active representatives among other gaps to ensure the
  gap invariant is restored.
\item
  Verify if \(s_{\#} = 2s_{\#}'\). If this is the case, the structure
  has reached its maximum capacity, and requires a resize operation that
  expands the library so it can hold more active representatives. The resulting library structure \(L\) is made up of twice the gaps
\(L_{1}\ldots L_{s_{\#}'}\) where each gap \(L_{i}\) contains a single
active representative \(R_{i}'\) and \(7\) buffer representatives. The process is as follows:
  \begin{enumerate}
  \def\labelenumii{\roman{enumii}.}
  \item
    Linearly scan through \(L_{1}\ldots L_{s_{\#}'}\) for the active
    representatives \(R_{1}'\ldots R_{s_{\#}}'\), exchanging them to a
    contiguous block at the front of \(D\). The scanning process begins
    at the \(\mathcal{\ell}_{2}\) of each gap as \(\mathcal{\ell}_{1}\) is
    guaranteed to be active by the gap invariant and linearly scans
    rightwards until it meets a buffer representative, exchanging the
    found active representatives in the process.
  \item
    Transport all active representatives back into \(L\), where
    \(R_{i}'\) is exchanged with the buffer representative at position
    \(8(i - 1)\) in \(L\), placing \(R_{i}'\) at the front of \(L_{i}\).
  \item
    Lastly, update \(s_{\#}' \coloneqq s_{\#} = 2s_{\#}'\) to account for the
    resize.
  \end{enumerate}
\end{enumerate}

\noindent
After insertion, the global order of active representatives is
preserved. The initial search step finds the last active representative
\(\leq x\) by \(R.\text{search}(x)\), then \(x\) is placed in the position
directly after that position, ensuring the new active representative
\(R_{z}'\) is constructed in its correct sorted position, obeying
\(R_{z - 1}' \leq R_{z}' < R_{z + 1}'\). The resize operation collects
all active representatives in the rebuild buffer and then places each
\(R_{i}'\) at position \(8(i - 1)\). Since the sequence \(8(i - 1)\) is
monotone increasing with \(i\), the active representatives appear in the
same order as they were collected, maintaining the global sorted
sequence \(R_{1}'\ldots R_{s_{\#}}'\).

Let us now derive the standard cost of \(R.\text{insert}(x)\) excluding the
cost of dynamic rebalancing. The first step requires at most
\(\lg s_{\#}' + 3\) comparisons by \hyperref[LAp1]{Lemma A.1}. The
second step requires exactly \(\lg 8 = 3\) comparisons as \(8\) is a
power of 2. The third step exchanges \(k - j + 1 \leq 7\) active
representatives, which requires \(O(\lg m)\) comparisons and
\(O(\lg m)\) moves. Constructing the representative in step 4
requires \(O(\lg m)\) moves. Verifying if \(k < 8\) was violated in step
5 requires only \(O(1)\) arithmetic operations.

The resize in step 6 is triggered following \(s_{\#}'\) insertions into
the structure since the last resize. The initial scanning phase makes
\(k\) comparisons and \(k\) exchanges for a gap containing \(k\) active
representatives, thus in total it makes exactly \(s_{\#}\) comparisons
and representative exchanges. Then, each active representative is
exchanged back into the library. Spread among \(s_{\#}'\) insertions,
using \(s_{\#} = 2s_{\#}'\), this is \(2\) comparisons and \(4\)
representative exchanges per insertion, which are \(O(\lg m)\)
comparisons and \(O(\lg m)\) moves. Following the resize, \(s_{\#}'\)
has doubled, and the structure requires \(s_{\#}'\) subsequent
insertions to trigger another resize, which has equal amortized cost by
the same argument.

\subsection{Rebalancing}
We proceed to consider the process of dynamic rebalancing. Recall that
we have an imaginary segment tree over the gaps
\(L_{1}\ldots L_{s_{\#}'}\) with nodes \(n_{1}\ldots n_{2s_{\#}' - 1}\),
where \(n_{1}\) is the root node covering all gaps, and leaf nodes
\(n_{s_{\#}'}\ldots n_{2s_{\#}' - 1}\) correspond with individual gaps.
For a node \(n_{k}\) covering the range \(L_{\ell}\ldots L_{r}\),
\(n_{2k}\) is its left child covering the left half of the range
\(L_{\ell}\ldots L_{\left\lfloor (l + r)/2 \right\rfloor}\), and
\(n_{2k + 1}\) is its right child covering the right half of the range
\(L_{\left\lfloor (l + r)/2 \right\rfloor + 1}\ldots L_{r}\).
This implies the sibling of a node \(n_{k}\) is \(n_{k \oplus 1}\) and
its parent is \(n_{\left\lfloor k/2 \right\rfloor}\) covering the
union of the ranges of its children \(n_{k}\) and \(n_{k \oplus 1}\). As
\(s_{\#}'\) is a power of 2, the general formula for the range covered
by a node \(n_{k}\) is \(L_{\ell}\ldots L_{r}\) where
\(l = \left( k - 2^{\left\lfloor \lg k \right\rfloor} \right)2^{\lg s_{\#}' - \left\lfloor \lg k \right\rfloor} + 1\)
and \(r = l + 2^{\lg s_{\#}' - \left\lfloor \lg k \right\rfloor} - 1\).
For a node \(n_{k}\) covering the range \(L_{\ell}\ldots L_{r}\), we define
\(\text{size}\left( n_{k} \right)\) as the number of active representatives in
the range covered by the node, that is the number of active
representatives in \(L_{\ell}\ldots L_{r}\), and
\(\text{span}\left( n_{k} \right)\) as the length of the range covered by the
node, which is given by \(r - l + 1\) or
\(2^{\lg s_{\#}' - \left\lfloor \lg k \right\rfloor}\) from the general
formulas of \(l\) and \(r\). As a parent covers the union of the ranges
of its children,
\(\text{size}\left( n_{k} \right) = \text{size}\left( n_{2k} \right) + \text{size}\left( n_{2k + 1} \right)\)
and
\(\text{span}\left( n_{k} \right) = \text{span}\left( n_{2k} \right) + \text{span}\left( n_{2k + 1} \right)\).
Additionally, since each gap contains at least \(1\) active
representative by the gap invariant,
\(\text{size}\left( n_{k} \right) \geq \text{span}\left( n_{k} \right)\). This
structure is not explicitly stored by the algorithm, but we logically
traverse the tree during dynamic rebalancing.

The key realization is that density-based balancing, e.g. in
\cite{FG05} are inherently limited to logarithmic
slack, incurring an \(O(\lg{\lg s_{\#}'})\) comparisons overhead per
search. Instead, we lazily keep nodes weight balanced, like scapegoat
trees in \cite{GR93}. For our structure, the invariant
for some internal node \(n_{k}\) is
 \begin{equation}
\left| \text{size}\left( n_{2k} \right) - \text{size}\left( n_{2k + 1} \right) \right| \leq \max\left( 1,\text{size}\left( n_{k} \right)/\lg s_{\#}' \right)
\end{equation}

\noindent
We let \(n_\text{leaf}\) be some leaf node in the tree whose ancestors all
satisfy (A.1), and \(n_{i}\) be the shallowest ancestor of \(n_\text{leaf}\)
such that \(\text{size}\left( n_{i} \right) < \lg s_{\#}'\). Using
\(\text{size}\left( n_{2k} \right) + \text{size}\left( n_{2k + 1} \right) = \text{size}\left( n_{k} \right)\),
we bound the size of the children of a node \(n_{k}\) that follows (A.1)
when \(\text{size}\left( n_{k} \right) \geq \lg s_{\#}'\)
\[\max\left( \text{size}\left( n_{2k} \right),\text{size}\left( n_{2k + 1} \right) \right) \leq \left( 1 + 1/\lg s_{\#}' \right)\text{size}\left( n_{k} \right)/2\]

\noindent
By the above relationship, the size of a node reduces by a factor of at
least \(\left( 1 + 1/\lg s_{\#}' \right)/2\) per level of
the tree given the size of the parent exceeds \(\lg s_{\#}'\), which is
true for all \(\left\lfloor \lg i \right\rfloor\) ancestors of \(n_{i}\)
as\(\ \)sizes of nodes moving up the tree are monotone increasing. Thus,
we bound \(\text{size}\left( n_{i} \right)\)
\begin{align*}
\text{size}(n_{i}) &\leq \left( (1 + 1/\lg s_{\#}')/2 \right)^{\lfloor \lg i \rfloor} s_{\#} \\
&\leq (1 + 1/\lg s_{\#}')^{\lfloor \lg i \rfloor} s_{\#}/2^{\lfloor \lg i \rfloor}
\end{align*}

\noindent
The expression \(\left( 1 + 1/u \right)^{u}\) approaches \(e\)
from below, and since \(i \leq 2s_{\#}' - 1\) by the definition of the
node indexes and \(s_{\#}'\) is a power of 2, we can deduce
\(\left\lfloor \lg i \right\rfloor \leq \lg s_{\#}\); thus
\(\left( 1 + 1/\lg s_{\#}' \right)^{\left\lfloor \lg i \right\rfloor} \leq \left( 1 + 1/\lg s_{\#}' \right)^{\lg s_{\#}'} \leq e\),
giving
 \begin{equation}
\text{size}\left( n_{i} \right) \leq es_{\#}/2^{\left\lfloor \lg i \right\rfloor}
\end{equation}

\noindent
For nodes \(n_{k}\) following (A.1) where
\(\text{size}\left( n_{k} \right) < \lg s_{\#}'\),
nt\(\left| \text{size}\left( n_{2k} \right) - \text{size}\left( n_{2k + 1} \right) \right| \leq 1\) is enforced by the invaria, and as
\(\text{size}\left( n_{k} \right) = \text{size}\left( n_{2k} \right) + \text{size}\left( n_{2k + 1} \right)\),
this implies
\[\max\left( \text{size}\left( n_{2k} \right),\text{size}\left( n_{2k + 1} \right) \right) \leq \left( \text{size}\left( n_{k} \right) + 1 \right)/2\]

\noindent
Repeating this linear recurrence from \(n_{i}\) for the remaining
\(\lg s_{\#}' - \left\lfloor \lg i \right\rfloor\) layers of the tree
until we reach \(n_\text{leaf}\) yields by induction
\begin{align*}
\text{size}\left( n_\text{leaf} \right) &\leq \left( \text{size}\left( n_{i} \right) + \left( 2^{\lg s_{\#}' - \left\lfloor \lg i \right\rfloor} - 1 \right) \right)/2^{\lg s_{\#}' - \left\lfloor \lg i \right\rfloor} \\[0.5em]
&\leq \text{size}\left( n_{i} \right)/2^{\lg s_{\#}' - \left\lfloor \lg i \right\rfloor} + 1
\end{align*}

\noindent
Applying the size bounds for \(n_{i}\) in (A.2), we obtain
\begin{align*}
\text{size}\left( n_\text{leaf} \right) &\leq \left( es_{\#}/2^{\left\lfloor \lg i \right\rfloor} \right)/2^{\lg s_{\#}' - \left\lfloor \lg i \right\rfloor} + 1\\
&\leq es_{\#}/2^{\lg s_{\#}'} + 1
\end{align*}

\noindent
Using \(s_{\#} \leq 2s_{\#}'\) yields the final bound for
\(\text{size}\left( n_\text{leaf} \right)\)
\begin{equation}
\text{size}\left( n_\text{leaf} \right) \leq 2e + 1
\end{equation}

\noindent
This bound is critical in proving the existence of a scapegoat node,
shown shortly. We will first describe the procedure for dynamic
rebalancing on a full gap \({L}_{i}\):

Firstly, initialize the procedure. We use the variables: \(j\) to denote
the current node in the traversal is \(n_{j}\), \(z\) to denote the size
of the node \(n_{j}\), and variables \(z_{1}'\) and \(z_{2}'\) to denote
the size of the child nodes of \(n_{j}\), where \(z_{1}'\) and
\(z_{2}'\) can correspond to any child. We begin at the leaf node
\(n_{j}\) that corresponds to the full gap \(L_{i}\), so
\(j = i + s_{\#}' - 1\) and \(z = 8\).

Secondly, traverse to the parent of the current node \(n_{j}\). Set
\(z_{1}' \coloneqq z = \text{size}\left( n_{j} \right)\), and
\(z_{2}' \coloneqq \text{size}\left( n_{j \oplus 1} \right)\). To find
\(\text{size}\left( n_{j \oplus 1} \right)\), first set \(z_{2}' \coloneqq 0\), then
use the general formula to find the range \(L_{\ell}\ldots L_{r}\) covered
by \(n_{j \oplus 1}\), and iterate through all gaps in
\(L_{\ell}\ldots L_{r}\), where for each gap a linear scan from
\(\mathcal{\ell}_{2}\) counts the number of active representatives in the
gap and adds it to \(z_{2}'\) (similar to during resizes, but without
exchanging active representatives). This process requires
\(\text{size}\left( n_{j \oplus 1} \right) = z_{2}'\) comparisons. Lastly, set
\(j \coloneqq \left\lfloor j/2 \right\rfloor\), moving the current node
marker to the parent of the current node, then
\(\left\{ z_{1}',z_{2}' \right\} = \{ \text{size}\left( n_{2j} \right),\text{size}\left( n_{2j + 1} \right)\}\),
which are correctly the size of the two children of \(n_{j}\). As
\(n_{j}\) covers the union of the ranges of its children, calculate
\(z \coloneqq z_{1}' + z_{2}'\). This moves the current node to its parent node
whilst maintaining the correctness of all variables.

Thirdly, verify if the current node \(n_{j}\) obeys the requirement
 \begin{equation}
\left| \text{size}\left( n_{2j} \right) - \text{size}\left( n_{2j + 1} \right) \right| > \max\left( 1,\text{size}\left( n_{j} \right)/\lg s_{\#}' \right)
\end{equation}

\noindent
Here we do not need to recompute the sizes of the nodes, as they are
already computed and stored in \(z\), \(z_{1}'\) and \(z_{2}'\). As
\(\text{size}\left( n_{2j} \right)\) and \(\text{size}\left( n_{2j + 1} \right)\) are
interchangeable in (A.4), the lack of differentiation between the left
and right children of \(n_{j}\) in \(z_{1}'\) and \(z_{2}'\) does not
matter. If the requirements are met, \(n_{j}\) is the scapegoat node,
and we rebuild the subtree rooted at the \(n_{j}\), distributing active
representatives under \(n_{j}\) as evenly as possible, then terminate.
Otherwise, return to step 2 and continue traversing upwards the tree.

Given that \({L}_{i}\) is full, a scapegoat node that obeys these
requirements must exist. Condition (A.4) is the negation of the invariant
(A.1); a node that violates (A.1) satisfies (A.4). By (A.3), if all
ancestors of \({L}_{i}\) obey the invariant (A.1), the size of \(L_{i}\)
is at most \(2e + 1\). However, as \({L}_{i}\) is full with
\(8 > 2e + 1\) active representatives, we have a contradiction, and thus
there must exist an ancestor that disobeys the invariant (A.1) and thus
obeys (A.4).

\subsection{Rebuilding}
The process for rebuilding the subtree rooted at the scapegoat node
\(n_{j}\) covering the range \(L_{\ell}\ldots L_{r}\) with
\(z = \text{size}\left( n_{j} \right)\) involves the following:

Firstly, by a linear scan of \(L_{\ell}\ldots L_{r}\) identical to the scan
in resizing, exchange all active representatives to a contiguous block
\(a_{1}\ldots a_{z}\) at the front of \(D\). Now the region
\(L_{\ell}\ldots L_{r}\) is composed entirely of buffer representatives.

Secondly, replace the active representatives into \(L_{\ell}\ldots L_{r}\).
For all gaps \(L_{k} \in L_{\ell}\ldots L_{r}\) with the corresponding leaf
node \(n_{k'}\) (where \(k' = k + s_{\#}' - 1\)) under \(n_{j}\),
determine the active representatives \(a_{l'}\ldots a_{r'}\) that should
be in \(L_{k}\), and exchange them into the first \(r' - l' + 1\)
positions of \(L_{k}\). We cannot simply distribute them naively in the
manner of \cite{FG05} as the invariant (A.1) is based
off weight instead of density which may remain violated even after a
naive redistribution, thus we define \(a_{l'}\ldots a_{r'}\) for all
nodes in the subtree as follows:

The scapegoat node \(n_{j}\) trivially covers all active representatives
\(a_{1}\ldots a_{z}\). For a node with current range
\(a_{l'}\ldots a_{r'}\), distribute the left half
\(a_{l'}\ldots a_{\left\lfloor \left( l' + r' \right)/2 \right\rfloor}\)
to the left subtree and the right half
\(a_{\left\lfloor \left( l' + r' \right)/2 \right\rfloor + 1}\ldots a_{r'}\)
to the right subtree. Rearranging, the left subtree receives
\(\left\lfloor \left( r' - l' \right)/2 \right\rfloor + 1\)
active representatives and the right subtree
\(\left\lceil \left( r' - l' \right)/2 \right\rceil\) active
representatives. From the property
\(0 \leq \left\lfloor \left( r' - l' \right)/2 \right\rfloor + 1 - \left\lceil \left( r' - l' \right)/2 \right\rceil \leq 1\)
for all positive integers \(l \leq r\), for any node \(n_{k}\) in the
rebuilt subtree
\(0 \leq \text{size}\left( n_{2k} \right) - \text{size}\left( n_{2k + 1} \right) \leq 1\),
spreading active representatives as evenly as possible, which guarantees
the invariant (A.1) is upheld in the rebuilt subtree. To calculate the
suitable range for all leaf nodes, we are unable to use the recursive
method in \cite{GR93} as it requires a recursion stack
of \(O( \lg z )\) words. Instead, we utilize a more complex
method that requires a stack \(K\) of only \(O( \lg z )\)
bits.

Start by initializing \(l' = 1\), \(r' = z\) and \(j' = j\) to denote
the current node in the traversal is the scapegoat node \(n_{j}\) which
should have all the collected active representatives
\(a_{1}\ldots a_{z}\) under it. \(K\) is initially empty. We now
describe three move primitives to move the current node \(n_{j'}\) to
the left child, right child, and parent node whilst updating the
variables \(l'\) and \(r'\) correctly. We assume that \(j',\ l'\),
\(r'\) and \(K\) are stored in some scope such that they are accessible
in the procedures.

The procedure to move to the left child pushes
\(\left\lfloor \left( r' - l' \right)/2 \right\rfloor + 1 - \left\lceil \left( r' - l' \right)/2 \right\rceil\)
into \(K\), then sets \(r' \coloneqq \left\lfloor (l' + r')/2 \right\rfloor\)
and \(j' \coloneqq 2j'\).

The procedure to move to the right child pushes
\(\left\lfloor \left( r' - l' \right)/2 \right\rfloor + 1 - \left\lceil \left( r' - l' \right)/2 \right\rceil\)
into \(K\), then sets
\(l' \coloneqq \left\lfloor (l' + r')/2 \right\rfloor + 1\) and
\(j' \coloneqq 2j' + 1\).

The procedure to move to the parent first reads the top of \(K\) and
stores it in a variable \(d\), and pops the top of \(K\). Then, if
\(n_{j}'\) is a left child, which is the case when \(j'\) is even, set
\(r' \coloneqq r' + (r' - l' + 1 - d)\). Otherwise, \(n_{j}'\) is a right child
and set \(l' \coloneqq l' - (r' - l' + 1 + d)\). Lastly, set
\(j' \coloneqq \left\lfloor j'/2 \right\rfloor\) regardless of which side
\(n_{j}'\) is on.

In the stack, we store the difference in sizes of the two child nodes
for every ancestor of the current node \(n_{j'}\) up to the scapegoat
node \(n_{j}\). As
\(0 \leq \left\lfloor \left( r' - l' \right)/2 \right\rfloor + 1 - \left\lceil \left( r' - l' \right)/2 \right\rceil \leq 1\),
this value is either \(0\) or 1, which can be represented using one bit.
Since the height of the subtree rooted at \(n_{j}\) is
\(\lg s_{\#}' - \left\lfloor \lg j \right\rfloor \leq \lg m\), the
entire stack fits in a single word of at least \(\lg m\) bits.

Trivially, the procedures to move to the left child and right child give
us the correct values \(l'\) and \(r'\) as they simply follow the
definitions of \(l'\) and \(r'\). Using the stored difference in sizes
and the property that the left child is always larger as
\(\left\lfloor \left( r' - l' \right)/2 \right\rfloor + 1 \geq \left\lceil \left( r' - l' \right)/2 \right\rceil\),
we can determine the range of active representatives distributed to the
sibling node from the current node, and as the parent node contains the
union of its children, this also allows us to determine the range of
active representatives distributed to the parent node.

We now show that moving to the parent correctly reverses moving to a
child. We start by assuming the variables \(j' = j_{i}'\),
\(l' = l_{i}'\) and \(r' = r_{i}'\) are in some arbitrary valid
state. After we move to the left child, \(j' = 2j'_{i}\),
\(l' = l_{i}'\),
\(r' = l_{i}' + \left\lfloor \left( r_{i}' - l_{i}' \right)/2 \right\rfloor\),
and we have
\(\left\lfloor \left( r_{i}' - l_{i}' \right)/2 \right\rfloor + 1 - \left\lceil \left( r_{i}' - l_{i}' \right)/2 \right\rceil\)
at the top of the stack. Then, if we move back to the parent,
\(j' = \left\lfloor 2j_{i}'/2 \right\rfloor = j_{i}'\),
\(l' = l_{i}'\), and
\begin{align*}
r' = {} & l_{i}' + \left\lfloor \left( r_{i}' - l_{i}' \right)/2 \right\rfloor \\ &+ \left( l_{i}' + \left\lfloor \left( r_{i}' - l_{i}' \right)/2 \right\rfloor - l_{i}' + 1 - \left( \left\lfloor \left( r_{i}' - l_{i}' \right)/2 \right\rfloor + 1 - \left\lceil \left( r_{i}' - l_{i}' \right)/2 \right\rceil \right) \right)
\end{align*}

\noindent
Cancelling out complementary terms
\[r' = l_{i}' + \left\lfloor \left( r_{i}' - l_{i}' \right)/2 \right\rfloor + \left\lceil \left( r_{i}' - l_{i}' \right)/2 \right\rceil\]

\noindent
Using
\(\left\lfloor \left( r_{i}' - l_{i}' \right)/2 \right\rfloor + \left\lceil \left( r_{i}' - l_{i}' \right)/2 \right\rceil = r_{i}' - l_{i}'\)
for all \(r_{i}' - l_{i}'\mathbb{\in Z}\), we obtain
\[r' = r_{i}'\]

\noindent
\(K\) is also reverted to its initial state as the newly pushed value is
popped out when we move back to the parent. By moving to the right child
then back to the parent, the same equality \(j' = j_{i}'\),
\(l' = l_{i}'\), \(r' = r_{i}'\) and unchanged \(K\) holds. Thus,
moving to the parent correctly reverses moving to a child.

By induction, starting from the root of the subtree \(n_{j}\) with the
valid initialized ranges, any sequence of moving to the left child,
right child, or parent yields a valid state with the correct block of
active representatives \(a_{l'}\ldots a_{r'}\) for the current node as
long as the current node is a valid node in the subtree.

Using the above three move primitives, perform a standard in-order
traversal of the subtree rooted at the scapegoat node. Upon reaching a
leaf \(n_{j'}\), exchange the active representatives
\(a_{l'}\ldots a_{r'}\) that should be in \(n_{j'}\) into the first
\(r' - l' + 1\) positions of the leaf node \(n_{j'}\)'s corresponding
gap \(L_{j' - s_{\#}' + 1}\). We know that a node \(n_{j'}\) is a leaf
and has no more children if \(j' \geq s_{\#}'\) by the definition of the
tree.

After dynamic rebalancing, the gap invariant is restored. Before the insertion, all gaps followed the gap invariant and contained less than $7$ active representatives. The insertion altered the size of $L_i$ to $8$, triggering the rebalancing. Thus, viewing from the scapegoat node, the size of the child containing $L_i$ is bounded by\(z_{1}' \leq 7\left( \text{span}\left( n_{2j} \right) - 1 \right) + 8 = 7span\left( n_{2j} \right) + 1\), and the size of the other child is bounded by $z'_2 \leq 7span(n_{2j})$. By condition (A.4), we know the difference in sizes between the two children is at least $2$. If $z'_1>z'_2$, then $z'_2\leq z'_1-2=7span(n_{2j})-1$, thus the size of the rebuilt subtree is given by $z'_1+z'_2\leq2(7span(n_{2j}))=7span(n_j)$. In the latter case where $z'_2>z'_1$, $z'_1\leq z'_2-2=7span(n_{2j})-2$, and summing gives the bound $z'_1+z'_2\leq2(7span(n_{2j}))-2 \leq 7span(n_j)$. Thus, this inequality connecting $z=\text{size}(n_j)$ and $\text{span}(n_j)$ holds in either case.

Because \(\text{size}\left( n_{2k} \right) - \text{size}\left( n_{2k + 1} \right) \leq 1\) for all nodes in the rebuilt subtree, by induction it can
be shown that any two nodes \(n_{k_{1}}\) and \(n_{k_{2}}\) on the same
level in the rebuilt subtree obeys
\(\left| \text{size}\left( n_{k_{1}} \right) - \text{size}\left( n_{k_{2}} \right) \right| \leq 1\).
Using this property, the size of a leaf node \(n_\text{leaf}\) in the rebuilt
subtree satisfies
\[\text{size}\left( n_\text{leaf} \right) \leq \left\lceil \text{size}(n_{j})/\text{span}(n_{j}) \right\rceil\]

\noindent
Shown above \(\text{size}\left( n_{j} \right) \leq 7span\left( n_{j} \right)\), so
\[\text{size}\left( n_\text{leaf} \right) \leq 7\]

\noindent
Likewise, the size of each leaf node is at least \(\left\lfloor \text{size}(n_{j})/\text{span}(n_{j}) \right\rfloor \geq 1\) by \(\text{size}\left( n_{j} \right) \geq \text{span}(n_{j})\). Thus, the gap invariant $0<k<8$ is restored.

We also show that the rebuild does not change the order of active
representatives. The linear scan collects active representatives from
the subtree into the rebuild buffer in the same order as they appear in
the library. At every step of the traversal downwards to the leaves, we
distribute the left half of the active representatives to the left
child, and the right half to the right child. By induction, for any two
collected active representatives \(a_{x}\) and \(a_{y}\) where
\(x < y\), \(a_{x}\) will be placed before \(a_{y}\) in the library,
maintaining local order within the rebuilt subtree. Furthermore, as
elements are only rearranged within the rebuilt subtree, global order is
also maintained.

Now, let us evaluate the total cost of an instance of dynamic
rebalancing.

At any step of the traversal, we set \(z_{1}' = z\), count
\(z_{2}' = \text{size}\left( n_{j \oplus 1} \right)\) using \(z_{2}'\)
comparisons, then set \(z = z_{1}' + z_{2}'\), increasing \(z\) by
\(z_{2}'\). Thus, it requires \(1\) comparison to count every additional
active representative. Initially, the count starts at \(8\), thus we
counted \(z - 8\) additional active representatives by this method,
which requires \(z - 8\) comparisons in total. Arithmetic operations
during traversal are negligible at \(O(1)\) per level traversed.

Then, the rebuild process first uses \(z\) comparisons and \(z\)
representative exchanges to move all active representatives to \(D\).
After which, they are replaced back into the library. This process
exchanges all \(z\) representatives back into the library in their
calculated positions using one exchange. By the standard analysis of
traversing a binary tree in in-order, the move procedures are called
\(O(z)\) times, and each use requires \(O(1)\) arithmetic operations,
totaling \(O(z)\) arithmetic operations. Summing up, dynamic rebalancing
makes \(O( z\lg m )\) comparisons, \(O( z\lg m )\)
moves, and \(O( z\lg m )\) arithmetic operations.

As the traversal phase does not make any moves, it cannot change the
order of active representatives, and as rebuilds do not change the order
of active representatives, active representatives remain in sorted order
after dynamic rebalancing.

\subsection{Amortized Analysis}
Although individual instances of dynamic rebalancing can cost up to
\(O( s_{\#}\lg m )\) when the root is rebuilt, as large
rebuilds are rare, the amortized cost of dynamic rebalancing is small.
While the main text provides an intuitive bound using a simplified accounting method, here we formalize this proof via a potential argument.

For an internal node \(n_{k}\), let the imbalance of \(n_{k}\) be
defined as
 \begin{equation}
\psi\left( n_{k} \right) = \max\left( 0,\left| \text{size}\left( n_{2k} \right) - \text{size}\left( n_{2k + 1} \right) \right| - 1 \right)
\end{equation}

\noindent
This allows us to define our non-negative potential function for some scaling constant $c$:
 \begin{equation}
\Phi = \sum_{i=1}^{s'_\#-1} c \cdot \psi(n_i) \lg^2 s'_\#.
\end{equation}

\noindent
When a new active representative is inserted into a gap \(L_{i}\), the
size of that leaf node and all its ancestors increases by \(1\). For each of the $\lg s'_\#$ ancestors of \(L_{i}\), exactly one child's size increases by \(1\), which may cause its imbalance to increase by \(1\). This, in turn may increase potential by $c \lg s'_\# \lg^2 s'_\#=c\lg^3 s'_\#$. Recall previously that the baseline actual cost of insertion, i.e. excluding dynamic rebalancing is $O(\lg m)$. By definition, the amortized cost is given by the actual cost of the operation plus the change in the potential function ($\Delta \Phi$). Thus, the aggregate cost of insertion prior to the rebuild is given by:
 \begin{equation*}
O(\lg m)+c\lg^3 s'_\#=O(\lg^3m).
\end{equation*}

\noindent
When a rebuild occurs, the found scapegoat node $n_{j}$ has $\psi\left( n_{j} \right) = \Omega(z/\lg s_{\#}')$ by condition (A.4). After rebuilding the subtree rooted in $n_{j}$, all internal nodes $n_{k}$ within the rebuilt topology obey $\left| \text{size}\left( n_{2k} \right) - \text{size}\left( n_{2k + 1} \right) \right| \leq 1$, strictly reducing the internal potential of the subtree to 0. Because the total size of the subtree rooted at $n_{j}$ is strictly conserved during the local structural change, the sizes and imbalances of all ancestors of $n_{j}$ remain completely unaffected. Since all node potentials are non-negative, the total drop in the global potential $\Phi$ is bounded below by the drop at $n_{j}$ itself, which is $c\cdot\psi\left( n_{j} \right) \lg^2 s'_\# = c \cdot \Omega(z \lg s'_\#) = c \cdot \Omega(z \lg m)$. By selecting the scaling constant $c$ to be sufficiently large, the net decrease in $\Phi$ absorbs the $O(z \lg m)$ actual cost of the rebuild. Thus, the rebuild step contributes nothing (or a negative amount) to the aggregate amortized cost. This means the total amortized cost of insertion, including any dynamic rebalancing it may trigger, is strictly bounded by the baseline cost.

Resizes cannot increase potential as they re-initialize all gaps uniformly with \(1\) active representative per gap which represents a perfectly balanced state where potential is 0. Thus, the aggregate cost stands.

\begin{lemma} \(R.\text{insert}(x)\) requires
\(O( \lg^{3}m )\) comparisons and
\(O( \lg^{3}m )\) moves amortized.
\end{lemma}

\section{Full Small Sort
Procedure}\label{appendix-b-full-small-sort-procedure}

Recall that the \emph{small sort} is designed to sort small subsequences
\(A'\) of
 \begin{equation}
m' = O(\lg^c m)\
\end{equation}

\noindent
active elements given a merge buffer \(B'\) of \(m'\) buffer
elements, alongside the hole and \(O(1)\) auxiliary words of \(w \geq \lg m\)
bits, where $c$ is an arbitrarily large, but constant value. This is achieved by a \(k\)-way merge sort, where
 \begin{equation}
k = 2^{\left\lceil \lg\left( w/\lg{\lg m} \right) \right\rceil} \geq \lg m/\lg{\lg m}
\end{equation}

\noindent
Let the number of merge passes \(k_{\#}\) be
 \begin{equation}
^{}k_{\#} = \left\lceil \lg m'/\lg k \right\rceil - 1
\end{equation}

\noindent
This is equivalent to the ceiling of the base-\(k\) logarithm of \(m'\)
minus \(1\). Lastly, we define the sequence \(r_{0}\ldots r_{k_{\#}}\)
where\(_{}\)
 \begin{equation}
r_{i} = k^{i + 1}
\end{equation}

\noindent
The algorithm begins with initial run sorting using table sort. We split
\(A'\) into runs \(A_{1}\ldots A_{\left\lceil m'/r_{0} \right\rceil}\)
where \(A_{i}\) is a contiguous block with size \(r_{0}\), and \(B'\)
into runs \(B_{1}\ldots B_{\left\lceil m'/r_{0} \right\rceil}\) by the
same definition. The final runs
\(A_{\left\lceil m'/r_{0} \right\rceil}\) and
\(B_{\left\lceil m'/r_{0} \right\rceil}\) may be short, containing less
than \(r_{0}\) elements. Processing all runs sequentially, for the
\(i\)th run execute the following:
\begin{enumerate}
\def\labelenumi{\arabic{enumi}.}
\item
  Using table sort in \cite{Knu98}, we create an array of auxiliary
  pointers to each of elements in a run \(A_{i}\), and sort the pointers
  with merge sort. As only pointers are moved, this process makes no
  data moves, only arithmetic operations.
\item
  Since the final destinations of all elements are now stored in the
  pointers, transport all elements to their destination using at most
  \(3\) moves per element. If \(k_{\#}\) is odd, transport all elements
  in sorted order into \(B_{i}\): iterating through the sorted pointer
  sequence, swap the element pointed to by the \(i\)th pointer with the
  \(i\)th element in \(B_{i}\); then logically switch the roles of
  \(A'\) and \(B'\), as \(B'\) now contains the active elements, and
  \(A'\) the buffer elements. Otherwise, the elements are permuted in-place into the sorted sequence. As active elements stay in \(A'\),
  there is no need to do a logical role switch in this case.
\end{enumerate}

\noindent
Every pointer to an element in a range of \(r_{0}\) elements requires
\(O(\lg r_{0})\) bits. By merge sort's linear space complexity, the
pointer array takes up \(O(r_{0})O( \lg r_{0} ) = O(w)\) bits
by (B.1) and (B.2), which fits in \(O(1)\) words. Each run of size \(k\)
is sorted with \(k\lg k - k + 1\) comparisons (by merge sort's
worst-case for power of 2 lengths), \(3k\) moves and \(O(k\lg k){}{}\)
arithmetic operations. The cost of this initial run sorting phase is
then \(m'\lg k - m' + o(m'){}\) comparisons and \(3m'\) moves.

After sorting the initial runs of size \(r_{0}\), we begin the process
of merging groups of \(k\) runs until \(A'\) is a contiguous sorted
block. Let \(i\) denote the current pass of merging. Iterating through
\(i = 1\ldots k_{\#}\), we execute the following process:

In the \(i\)th pass of merging, \(A'\) consists of runs of size
\(r_{i - 1}\), denoted by
\(A_{1}\ldots A_{\left\lceil m'/r_{i - 1} \right\rceil}\), where the
final run \(A_{\left\lceil m'/r_{i - 1} \right\rceil}\) may be short,
containing less than \(r_{i - 1}\) elements. Note that \(A'\) is not
necessarily the original \(A'\), but the current logical \(A'\) that may
be the original \(B'\), and similarly \(B'\) is not necessarily the
original \(B'\). We divide \(B'\) into runs
\(B_{1}\ldots B_{\left\lceil m'/r_{i} \right\rceil}\) analogously, but
with size \(r_{i}\). Now, we define groups of \(k\) runs from \(A'\),
where the \(j\)th group contains runs \(A_{k(j - 1) + 1}\ldots A_{kj}\).
Iterating through groups
\(j = 1\ldots\left\lceil m'/r_{i} \right\rceil\) sequentially, the
\(j\)th group is processed as follows:
\begin{enumerate}
\def\labelenumi{\arabic{enumi}.}
\item
  Construct a selection tree on pointers \(a_{1}\ldots a_{k}\) to the
  first elements of the \(k\) runs \(A_{k(j - 1) + 1}\ldots A_{kj}\) to
  select the minimum (rank of pointers dependent on elements they point
  to). Alongside each pointer to the first elements of the runs, we also
  store a pointer to the position right after end of each run, denoted
  by \(a_{1}'\ldots a_{k}'\). Lastly, we have a pointer \(b\)
  initialized to point to the beginning of the run \(B_{j}\).
\item
  Use the selection tree to select the minimum \(a_{x}\) from the
  current front of all runs (\(a_{1}\ldots a_{k}\)).
\item
  Swap \(a_{x}\) with \(b\), then increment \(a_{x}\) and \(b\) so they
  point to the next active element in \(A_{(j - 1)k + x}\) and the next
  available space in \(B_{j}\) respectively.
\item
  Repair the selection tree to select the next merged element. For runs
  where \(a_{x} = a_{x}'\), the run is exhausted of active elements, and
  we treat them as \(\infty\) sentinels.
\item
  Repeat steps 2-5 until the root becomes an \(\infty\) sentinel.
\end{enumerate}

\noindent
This process merges the group of runs \(A_{k(j - 1) + 1}\ldots A_{kj}\)
into one contiguous sorted run stored in \(B_{j}\), and
\(A_{k(j - 1) + 1}\ldots A_{kj}\) now hold only buffer elements. For the
last potentially short group, we use additional logic to calculate the
number of runs, then use the same merge procedure.

After processing all groups, logically swap roles of \(A'\) and \(B'\),
as all active elements have been swapped over to the other run. Now, if
\(r_{i} \geq m'\), which is the case when \(i \geq k_{\#}\), all \(m'\)
active elements have been merged into a sorted run. Otherwise, continue
to next pass of merging. When the procedure terminates, the logical
active region containing all active elements is guaranteed to be the
original active region due to the parity handling in the initial run
sorting step. This is because there are \(k_{\#}\) merge passes, and
every merge pass swaps all active elements to the opposite region. If
\(k_{\#}\) was even, all active elements would end up in the region they
started, which is the original \(A'\) as we would have used the cycle
method that kept active elements in \(A'\). And if \(k_{\#}\) was odd,
all active elements would end up in the opposite region, which is the
original \(A'\) as we would have exchanged them to \(B'\) in the initial
run sorting phase.

We now analyze the cost of a single pass of merging. The first
initialization step of merging a group of runs constructs a selection
tree over \(k\) pointers, requiring \(O(k)\) comparisons and \(O(k)\)
arithmetic operations, then uses \(O(1)\) arithmetic operations to
initialize \(b\), which costs \(o(m)\) over all groups. Every pointer to
an element from a range of \(m'\) elements requires
\(O(\lg{m'}) = O( \lg{\lg m} ){}\) bits by (B.1). As the
selection tree on \(\leq k\) entries has \(O(k)\) nodes, the required
number of bits for the selection tree is
\(2O(k)O( \lg{\lg m} ) = O(w)\) by (B.2), which fits in
\(O(1)\) words.

Subsequently, every merged element is selected from the selection tree
using exactly \(\lg k\) comparisons as \(k\) is a power of 2 by (B.2) and
exchanged into the next available position in \(B'\) using \(3\) moves.
The arithmetic operations made in assigning and dereferencing pointers
are\(\) \(O(\lg k)\). For the last short group, the selection tree may
contain less than \(k\) entries, but in any case, the number of
comparisons to select each element do not exceed \(\lg k\).

Thus, the total cost of a merge pass is \(m'\lg k + O( m' )\)
comparisons and \(3m'\) moves, with \(O(m'\lg k)\) arithmetic
operations. \(\)A naive bound for the total cost of the entire
algorithm, using the above bound for every pass would give
\(m'\lg k - m' + k_{\#}\left( m'\lg k + O( m' ) \right) = m'\lg m' + O(m'\lg{\lg{m'}})\)
comparisons and \(3(k_{\#} + 1)m'\) moves by (B.2) and (B.3). However, we
can obtain a tighter bound by carefully accounting for the comparisons made in the final merge pass.

When we conduct the final merging pass, \(A'\) is made up of runs of
size \(r_{k_{\#} - 1}\) from the previous merging passes, with the last
run being potentially short. Thus, there are at most
\(k' = \left\lceil m'/r_{k_{\#} - 1} \right\rceil=\lceil m'/k^{k_\#}\rceil\) runs. Selecting each
element from a selection tree of \(k'\) entries require only
\(\left\lceil \lg{k'} \right\rceil\) comparisons. For \(m'\) elements,
this is \(m'\left\lceil \lg{k'} \right\rceil + O( m' ){}\)
comparisons. Applying the naive bound for the first \(k_{\#} - 1\)
passes, and the tighter bound for the final pass, we have the total
number of comparisons \(C''(m')\) to sort a sequence of \(m'\) elements
using the algorithm
\begin{align*}
C^{''}\left( m' \right) &= m'\lg k - m' + (k_{\#} - 1)\left( m'\lg k + O( m' ) \right) + m'\left\lceil \lg{k'} \right\rceil + O( m' ) \\
&\leq m'\lg k^{k_{\#}} + m'\left\lceil \lg\left\lceil m'/k^{k_{\#}} \right\rceil \right\rceil - m' + o(m') \\
&\leq m'\lg k^{k_{\#}} + m'\lg(m/k^{k_{\#}}) + m' - m' + o(m') \\
&\leq m'\lg m' + o( m' )
\end{align*}

\noindent
The number of moves made by the final merge pass is still \(3m'\). As a
result, the naive bound is tight, which counts \(3m'\) moves for the
initial run sorting, and \(3m'\) moves for each of the \(k_{\#}\)
subsequent merge passes. By (B.1), (B.2) and (B.3),
\(k_{\#} = \left\lceil \lg m'/\lg k \right\rceil - 1 = O(1)\) when $m'=O(\lg^c m)$ and
word width \(w \geq \lg m\) bits, giving the final bounded moves
\(M''\left( m' \right) = 3\left( k_{\#} + 1 \right)m' = O(m')\).

\begin{lemma} For any constant $c$, a sequence of
\(m' = O(\lg^c m )\) active elements can be
sorted using no more than \(m'\lg{m'} + o(m')\) comparisons and
\(O(m')\) moves if given a merge buffer of \(m'\)
buffer elements.
\end{lemma}

\section{Analysis of Cache Lines}\label{appendix-c-analysis-of-cache-lines}

Recall that we have two designs of cache lines, the randomized
and deterministic design.

\subsection{Randomized Cache Lines}
First, we analyze randomized cache lines. We have
 \begin{equation}
g = 2^{\left\lceil \lg{\lg m} \right\rceil}
\end{equation}

\noindent
A randomized cache line contains at most \(g\) active elements spread in
\(2g\) spaces, then the probability of any uniformly random \(g_{i}\)
being active is at most \(1/2\). Thus, from uniformly randomly probing
in step 1 of insertion, the expected number of probes is bounded by
\(1/(1 - 1/2) = 2\). This is averaged over all possible streams of
random bits (``coin flips'') used to choose random \(g_{i}\). Each probe
uses a single comparison to check if \(g_{i}\) is buffer, giving \(2\)
comparisons on average for probing in step 1. Then, swapping \(x\) with
the found buffer space \(g_{i}\) in step 2 requires 3 data moves.
Incrementing \(d\) in step 3 is implemented as a binary increment over
its \(\left\lfloor \lg g \right\rfloor + 1\) bits. A single increment
may read and flip all bits, but over a sequence of an arbitrary number
of increments from \(0\), the amortized cost of each increment
is\(\ O(1)\). This is because the \(j\)th bit is read then flipped every
\(2^{j}\) increments, and by aggregate analysis over an arbitrary number
\(k\) increments from \(0\), we have the amortized number of read and
write operations per increment
\[\left( \sum_{i = 0}^{k}\left\lfloor 1/2^{i} \right\rfloor \right)/k < 2\]

\noindent
This bound holds for any positive integer \(k\). As every read operation
requires \(1\) comparison and every write \(3\) moves, this is \(2\)
comparisons and \(6\) moves amortized. The final step, checking if the
cache line is full requires a read operation of the bit \(d_{\lg g}\).
This is because if bit \(d_{\lg g}\) (this is well defined as \(g\) is a
power of 2 by (C.1)) with priority \(2^{\lg g} = g\) by is 1,
\(d \geq g\), and as cache lines are maintained to not hold more than
\(g\) elements, \(d \leq g\), thus \(d = g\), and the cache line is full
since it contains the maximum amount of \(g\) active elements. In total,
inserting into randomized cache requires \(5\) comparisons expected and
\(9\) moves amortized.

\begin{lemma} Inserting into a randomized cache
line requires \(O(1)\) comparisons on average and \(O(1)\)
moves amortized.
\end{lemma}

\noindent
Let $p$ denote the number of failed probes. We now show that the probability for \(p > 2m\) is exponentially small in \(m\). Observe that though probes are not independent, every
individual probe has probability at least \(1/2\) for success
regardless, and the randomized sort finishes successfully when it
accumulates \(m\) such successes. Thus, the behavior of the number of
failed probes \(p\) when the randomized sort finishes is stochastically dominated by a negative binomial distribution for \(m\) successes with success probability at
least \(1/2\). Consequently,
\[\Pr\lbrack p > 2m\rbrack \leq \Pr\lbrack NB(m,1/2) > 2m\rbrack\]

\noindent
Applying the Chernoff bound and optimizing the rate function yields the tail bound
 \begin{equation}
\Pr\lbrack p > 2m\rbrack \leq (27/32)^{m}
\end{equation}

\noindent
To connect this tail bound to the overall comparison count, we separate the cost of probing from each cache insert: the remaining operations are $O(1)$ amortized, equating to $O(m)$ for $m$ cache inserts. For the $i$th cache insert, let there be $p_i$ failed probes before the final, successful probe, totaling $p_i+1$ comparisons. Because $\sum p_i=p$, the cost of probing is $\sum p_i+1 \leq p+m$. Combining with \hyperref[L3]{Lemma 2.4} and \hyperref[L4]{Lemma 2.5}, the total computational cost is $m \lg m + O(m) + p$ comparisons and $O(m)$ moves. Thus, from probability bound in (C.2), the probability the algorithm makes $>m \lg m + O(m) + 2m=m \lg m+O(m)$ comparisons is $\leq(27/32)^m$. Note that as the expected number of failed probes is $m$, this bound represents the probability of $>m$ more comparisons over the expected amount.

\subsection{Deterministic Cache Lines}
Now, we turn to deterministic cache lines. We will restate the
definition of the sequence \(g_{0}\ldots g_{t' + 1}\) with base cases
\begin{gather}
g_{0} = 1 \\
g_{1} = 2^{\left\lceil \lg^{(t)}m \right\rceil}
\end{gather}

\noindent
and for all \(2 \leq i \leq t'\) follows the recursive relationship
 \begin{equation}
g_{i} = g_{i - 1}2^{g_{i - 1}/2^{i - 1}}
\end{equation}

\noindent
By a straightforward induction on the recurrence, \(g_{i}\) is a power
of 2 for all \(0 \leq i \leq t'\). We define \(t'\) as the largest
positive integer such that
 \begin{equation}
g_{t'} < \lg m
\end{equation}

\noindent
To account for any remaining space, we have the final term (overrides (C.4) when $t'=0$)
\begin{equation}
g = g_{t' + 1} = g_{t'}\left\lceil \lg m/g_{t'} \right\rceil \leq g_{t'}2^{g_{t'}/2^{t' - 1}}
\end{equation}

\noindent
We bound \(g\) by the definition of \(t'\)
 \begin{equation*}
\lg m \leq g \leq \lg m + g_{t'} \leq 2\lg m
\end{equation*}

\noindent
Indeed, the definition of \(g\) satisfies (4). Then the size of each
cache line is
\[\sum_{i = 1}^{t' + 1}g_{i} \leq 2g\]

\noindent
For easy access to this sequence, we can store the entire sequence in
auxiliary memory by leveraging how the terms in the sequence do not
surpass \(g\) allowing us to store each term
\(\left\lfloor \lg g \right\rfloor + 1 = O( \lg{\lg m} )\)
bits. For \(t' + 1 \leq t \leq \lg^{*}m\) terms this is
\(O( \lg{\lg m} \cdot \lg^{*}m ){}\) bits, small enough to
fit in \(O(1)\) words as each word is at least \(\lg m\) bits. Note that
this sequence is only stored once for all cache lines as they have
identical internal structures, not individually stored for each cache
line which would require \(O(r_{\#})\) words.

We now analyze the cost of \(\text{insert}(x)\) for deterministic cache lines.
At every iteration, we move \(g_{i - 1}\) elements from level \(i - 1\)
to level \(i\) using
\(\left\lceil \lg{(g_{i}/g_{i - 1})} \right\rceil{}\) comparisons and
\(g_{i - 1}\) swaps which are \(3g_{i - 1}\) moves. As elements cannot
be moved to different levels in any other way, every element passes
through every level once. For the \(i\)th iteration, we calculate the
comparison cost per element spread over \(g_{i - 1}\) elements
\(C_{G}(i)\)
\[C_{G}(i) = \left\lceil \lg\left( g_{i}/g_{i - 1} \right) \right\rceil/g_{i - 1}\]

\noindent
By the recursive relationship of (C.5) and (C.7), we have the bounds for
\(i \geq 2\)
\begin{align*}
C_{G}(i) &\leq \lg 2^{g_{i - 1}/2^{i - 1}}/g_{i - 1} \\
&\leq 1/2^{i - 1}
\end{align*}

\noindent
Using base cases (C.3) and (C.4), we find the comparison cost of the first
iteration
\begin{align*}
C_{G}(1) &\leq \left\lceil \lg g_{1} \right\rceil/g_{0} \\
&\leq \left\lceil \lg^{(t)}m \right\rceil
\end{align*}

\noindent
Summing over all levels, we obtain the final comparison cost for inserting into a deterministic cache line
\begin{align*}
\sum_{i = 1}^{t' + 1}{C_G(i)} = C_G(1) + \sum_{i = 2}^{t' + 1}{C_G(i)} = \lceil \lg^{(t)} m \rceil + \sum_{i = 2}^{t' + 1}{1/2^{i-1}} < \lg^{(t)} m +2
\end{align*}

\noindent
For every cache block an element passes through, \(3\) moves are used to
swap the element from its previous cache block to its new position in
the cache block of the next level. Over \((t' + 1)\) cache blocks, the
moves for insertion are \(3(t' + 1)\). With
\(g_{1}' = \left\lceil \lg^{(t)}m \right\rceil\), and
\(g_{i}' = 2^{g_{i - 1}'}\), by a straightforward induction on the
recurrence in (C.5), one can show that for all \(1 \leq i \leq t'\)
 \begin{equation}
g_{i} > g_{i}'
\end{equation}

\noindent
We claim that \(t' \leq t - 1\). If it is not, we have \(t' \geq t\) as
\(t'\) must be an integer, then \(g_{t'} > g_{t'}' \geq g_{t}' = \lg m\) by
(C.8), but by definition \(g_{t'} < \lg m\) in (C.6), giving us a
contradiction. Thus, \(t' \leq t - 1\). This gives us the final bound
for inserting into a deterministic cache line of \(3t' \leq 3(t - 1 + 1) = 3t\)
moves.

\begin{lemma} Inserting into a deterministic cache
line requires \(\lg^{(t)}m + 2\) comparisons and \(3t\)
moves.
\end{lemma}

\begin{proof}[Proof of Remark 2.8]
A key advantage of deterministic cache lines is its stability on active elements. Here we formally prove this property by analyzing the relative order of equal active elements across the two distinct phases of the algorithm: insertion and extraction.

\textbf{Insertion.} We prove stability by induction on the sequence of inserted elements from the input array $A$. During initialization, the first $\lfloor s/2 \rfloor$ elements of $A$ are directly swapped in order into $S_0$, trivially preserving their relative order. For any subsequent element $x \in A$ processed chronologically, assume by induction that the current global logical sequence $S_0 f'_1 S_1 G'_1 f'_2 S_2 G'_2 \dots f'_{s_\#} S_{s_\#} G'_{s_\#}$ maintains the relative input order of all previously inserted equal active elements. To advance the inductive step, it suffices to establish that after inserting $x$ into the structure: (i) $x$ is placed strictly after all existing equal instances of $x$ in the logical sequence, and (ii) the relative order of all existing equal active elements remains unchanged. We verify both conditions by step-by-step inspection of the insertion procedure:
\begin{enumerate}
    \item The initial search query, $R.\text{search}(x)$ locates the active representative $R'_z$ satisfying $f'_z \le x < f'_{z+1}$, i.e. the last active representative $\leq x$. Consequently, no previously inserted elements equal to $x$ can exist at or after $f'_{z+1}$, due to the ordering invariant.
    \item The element is then inserted into the cache line $G'_z$. The cache insertion initially appends $x$ to back of the first level. As the first level physically occupies the end of a cache line, this places $x$ after all previous active elements in $G'_z$. Because no equal active elements exist at or beyond $f'_{z+1}$, this placement guarantees $x$ is positioned after all existing equal elements within the global sequence. Any subsequent element movement through deterministic cache levels does not change the order of active elements, as each lower level block is placed after all active sub cache blocks in the higher level, meaning the active elements in the moved cache block remains behind all active elements in upper levels, but before all active elements in the lower levels. 
    \item When $G'_z$ fills and is flushed, its active elements are swapped into the next available buffer spaces of $S_z$ in descending order of cache level, exactly as cache lines are physically arranged. This process appends the entire contents of $G'_z$ directly to the back of $S_z$. As $S_z$ directly precedes $G'_z$ in the logical sequence, the order of active elements is unchanged.
    \item If a flush causes $S_z$ to reach its maximum capacity, the splitting process explicitly uses stable selection and partitioning procedures to ensure the relative order of equal active elements remains constant. Allocating the new segment, moving the $\ge a_{\text{med}}$ block, and creating the new active representative simply redistributes the $s$ active elements originally in $S_z$ to occupy an expanded contiguous zone $S_z G'_z f'_{z+1} S_{z+1}$ (as shown in Figure 2) without permuting the elements. To confirm that $a_{\text{med}}$ assumes this exact position as $f'_{z+1}$, note that $R.\text{insert}$ internally utilizes $R.\text{search}$ to determine the insertion location. Because $f'_z \le a_{\text{med}} \le f'_{z+1}$ by the ordering invariant, $R.\text{search}$ is guaranteed to locate $f'_z$, constructing the new active representative containing $a_{\text{med}}$ directly after $f'_z$ to take the slot of $f'_{z+1}$. Any subsequent dynamic rebalancing or resizing operations within $R$ merely collect and redistribute active elements locally within a contiguous interval of gaps, preserving their relative order.
\end{enumerate}

\noindent
\textbf{Extraction.} During extraction, all remaining cache lines $G'_i$ are flushed into their corresponding segments $S_i$, which preserves the relative order of active elements as established previously. Each segment $S_i$ is then locally sorted using the small sort routine, making use of table sort and $k$-way merging, both of which are inherently stable, maintaining the relative order of active elements. Finally, the active elements are swapped back into $A$ in the order $S'_0 f'_1 S'_1 f'_2 \dots f'_{s_\#} S'_{s_\#}$ (all $G'_i$ contain no active elements and thus are safely omitted). Since the insertion phase maintains global relative order and each segment group is stably processed, this final extracted array retains the exact input order of equal elements.
\end{proof}

\section{In-Place Conversion}\label{appendix-d-in-place-conversion}

We now present the process for converting the algorithm that uses
additional buffers in Section 2 to the full in-place algorithm in
Section 3.

As stated in Section 3, the buffer memory is made up of buffer elements,
which are all greater or equal to a pivot \(p\),
whilst all active elements are less than $p$, allowing active and buffer elements to be differentiated via $1$ comparison against $p$.

The bit vector \(V\) is also formed by two blocks \(L\) and \(R\), where the pair of elements \((l_{i},r_{i})\) are used to
represent the \(i\)th bit. Initially, every element in \(L\) needs to be strictly less than every element in \(R\) (so no pair can be equal).

We now show how these regions are created, beginning with the bit
vector. Using heapselect, select the smallest and largest
\(v=n/\lg^{2}n\) elements and place them in contiguous blocks at the front
and back of \(\mathcal{A}\), changing the configuration of
\(\mathcal{A}\) to \(L\mathcal{A'}R\). We use a single comparison to
verify if the last element in \(L\) is strictly less than the first
element of \(R\). If so, the block \(\mathcal{A'}\) contains only equal
values, then the entire input array is already sorted, and the algorithm
terminates. Conversely, if the blocks \(L\) and \(R\) pass the test
above, they can be used to simulate a bit vector of \(n/\lg^{2}n\) bits.
As \(m \leq n\), the bit vector is sufficiently large for all instances
of sorting. This step requires \(O(n+v\lg n)=O(n)\) comparisons and moves.

Now, we repeatedly create active and buffer regions in a partition-based
loop. At any iteration, the configuration of \(A'\) is \(A_{S}A_{U}\),
where \(A_{S}\) contains elements in their correct sorted position and
\(A_{U}\) unsorted elements. Particularly, if \(A_{U}\) were sorted
using some algorithm, the entire block \(A'\) would be sorted, and \(A\)
would also be sorted as a whole. Initially, \(A_{S}\) is empty and
\(A_{U}\) contains the entire block \(A'\). Every iteration shrinks the
size of \(A_{U}\) by a factor of at least \(1/4\), until eventually
\(\left| A_{U} \right| \leq 2^{19} = 524288\), then the entire \(A_{U}\)
is sorted as a base case. The process for the \(i\)th iteration, where
\(n_{i} = |A_{U}|\), is as follows:
\begin{enumerate}
\def\labelenumi{\arabic{enumi}.}
\item
  By the in-place selection algorithm of
  \cite{GK06}, find the element \(p\) in \(A_{U}\)
  with rank \(\left\lceil n_{i}/4 \right\rceil\).
\item
  Partition \(A_{U}\) by pivot \(p\), such that the configuration of
  \(A_{U}\) becomes \(A_{<}pB_{\geq}\), where \(A_{<}\) contains
  elements \(< p\), and \(B_{\geq}\) contains elements \(\geq p\). We
  denote \(\left| A_{<} \right| = n_{i, <}\) and
  \(\left| B_{\geq} \right| = n_{i, \geq}\). As \(p\) has rank
  \(\left\lceil n_{i}/4 \right\rceil\),
  \(n_{i, <} \leq \left\lceil n_{i}/4 \right\rceil - 1 \leq n_{i}/4\),
  then
  \(n_{i, \geq} \geq n_{i} - \left\lceil n_{i}/4 \right\rceil \geq (3/4)n_{i} - 1\).
\item
  Sort the block \(A_{<}\) by the algorithm from Section 2, using the
  blocks \(L\) and \(R\) to simulate the bit vector \(V\), and
  \(B_{\geq}\) as the buffer memory. We can differentiate active
  elements originating from \(A_{<}\) with buffer elements originating
  from \(B_{\geq}\) by a single comparison against \(p\). As
  \(n_{i, \geq} \geq 3n_{i}/4 - 1 = 3n_{i, <} - 1\), \(B_{\geq}\) is
  indeed adequately large to function as the buffer memory to sort
  \(A_{<}\). This leaves \(A_{U}\) in the form \(A_{< ,S}pB_{\geq}'\),
  where \(A_{< ,S}\) is the sorted sequence \(A_{<}\) and \(B_{\geq}'\)
  is some permutation of \(B_{\geq}\).
\item
  Partitioning \(B_{\geq}'\) by pivot \(p\), we isolate all elements
  \(= p\) in \(B_{\geq}'\). The resulting structure of \(A_{U}\) is
  \(A_{< ,S}pPB_{>}\), where \(P\) contains \(n_{i, =}\) elements which
  are \(= p\) and \(B_{>}\) contains \(n_{i, >}\) elements that are
  \(> p\).
\item
  Notice that the elements \(A_{< ,S}pP\) are already in their correct
  sorted order, thus we update \(A_{S}\) to include \(A_{< ,S}pP\), and
  shrink \(A_{U}\) to only include \(B_{>}\), then start a new iteration
  with \(n_{i + 1} = |B_{>}|\). If \(n_{i + 1} \leq 2^{19}\), we sort
  the entire \(A_{U}\) by the method in
  \hyperref[handling-short-blocks]{Section 2.7} without using a buffer,
  then terminate. As \(p\) has rank
  \(\left\lceil n_{i}/4 \right\rceil\),
  \(\left| B_{>} \right| \leq n_{i} - \left\lceil n_{i}/4 \right\rceil \leq (3/4)n_{i}\).
\end{enumerate}

\noindent
We now calculate the cost of a single iteration. The in-place selection of \(p\)
requires \(O(n)\) comparisons and \(\varepsilon n\) moves by
\cite{GK06}. Then, partitioning \(A_{U}\) requires
\(n_{i}\) comparisons and \(2n_{i, <} + 1\) moves. Let \(C'(m)\) and
\(M'(m)\) denote the number of comparisons and moves respectively to
sort a block of size \(m\) with additional memory. Then, sorting
\(A_{<}\) using the buffer memories in step 3 requires \(C'(n_{i, <})\)
comparisons and \(M'(n_{i, <})\) moves. Lastly, partitioning
\(B_{\geq}\) to remove equal elements requires \(n_{i, \geq}\)
comparisons and \(2n_{i, =} + 1\) moves.

By straightforward induction, we bound \(n_{i} \leq (3/4)^{i}n\). With
\(I\) as the number of iterations, we have
 \begin{equation}
\sum_{i = 1}^{I}n_{i} \leq n\sum_{i = 1}^{I}{(3/4)^{i}\ } \leq n/(1 - 3/4) = 4n
\end{equation}

\noindent
As every iteration expands \(A_{S}\) by \(n_{i, <} + n_{i, =} + 1\)
until its size reaches \(\mathcal{|A'|}\) and the algorithm terminates,
then
 \begin{equation}
\sum_{i = 1}^{I}n_{i, <} \leq \left( \sum_{i = 1}^{I}{n_{i, <} + n_{i, =} + 1} \right)\mathcal{= |A'|} \leq n
\end{equation}

\noindent
Summing over all iterations, the number of comparisons \(C(n)\) to sort
the input array \(A\) is
\begin{align*}
C(n) &= O(n) + \left( \sum_{i = 1}^{I}{O( n_{i} ) + n_{i} + C'\left( n_{i, <} \right) + n_{i, \geq}} \right) \\
&= O(n) + \left( \sum_{i = 1}^{I}{C'\left( n_{i, <} \right) + O( n_{i} )} \right)
\end{align*}

\noindent
Applying (D.1), we obtain
\[C(n) = O(n) + \sum_{i = 1}^{I}{C'(n_{i, <})}\]

\noindent
Let $F(m)$ denote all terms in $C'(m)$ that grow super-linearly. From Theorems \hyperref[T6p1]{2.6} and \hyperref[T6p2]{2.7}, $F(m) = m \lg m$ or $F(m) = m \lg m + m \lg^{(t)}m$. Separating $C'(n_{i,<})$ gives:
\[C(n) = O(n) + \sum_{i = 1}^{I}\left( {F(n_{i, <})+O(n_{i,<})} \right) \]

\noindent
For both variations of $F(m)$, the expression $F(m)/m$ is non-decreasing when $F(m)$ is well defined. Therefore, $F(m)$ is superadditive. Using (D.2) and the superadditivity of $F(m)$:
\[C(n) \leq F(n)+O(n) = C'(n)+O(n). \]

\noindent
By a parallel derivation for the moves \(M(n)\) (note that $F(m)=0$ as $M'(m)$ grows linearly with $m$)
\begin{align*}
M(n) &= O(n) + \left( \sum_{i = 1}^{I}\varepsilon n_{i} + 2n_{i, <} + 1 + M'\left( n_{i, <} \right) + 2n_{i, =} + 1 \right) \\
&\leq M'(n) + O(n).
\end{align*}

\noindent
(Where \(\varepsilon > 0\) denotes an arbitrarily small, but fixed, real
constant.)

\begin{lemma}
The input array \(\mathcal{A}\)
can be sorted in-place using no more than \(C'(n) + O(n)\)
comparisons and \(M'(n) + O(n)\) moves given an algorithm
that sorts a sequence of \(m\) elements with additional memory
using \(C'(m)\) comparisons and \(M'(m)\) moves.
\end{lemma}

\noindent
Finally, we confirm that the exponentially high probability bound holds on $n$ for the in-place, randomized algorithm. By \hyperref[L3]{Lemma 2.4} and (D.2), across all instances there are a total of \(k \leq n\) cache inserts. Since failed probes accumulate monotonically with each insertion, we couple the actual execution with a hypothetical one that performs exactly $n$ insertions using the same random bits. Under this coupling, the actual number of failed probes $p$ is stochastically dominated by the number of failed probes in the $n$-insertion process. Paralleling the reasoning of
\hyperref[appendix-c-analysis-of-cache-lines]{{[}App. C{]}},
we therefore obtain the same tail bound on $p$:
\begin{align*}
\Pr\lbrack p > 2n\rbrack &\leq \Pr{\lbrack NB(n,1/2) > 2n\rbrack} \\
&\leq (27/32)^{n}
\end{align*}

\section{Notation Glossary}\label{appendix-e-notation-glossary}

In this glossary, notation is grouped by section. Each table provides the definition of relevant notation in the context of their respective sections.

\begin{center}

\textbf{Table E.1:} \hyperref[sorting-with-additional-memory]{Section 2} Notation
Table

{\def\LTcaptype{none} 
\begin{longtable}[]{@{}
  >{\raggedright\arraybackslash}p{(\linewidth - 2\tabcolsep) * \real{0.1141}}
  >{\raggedright\arraybackslash}p{(\linewidth - 2\tabcolsep) * \real{0.8859}}@{}}
\toprule\noalign{}
\begin{minipage}[b]{\linewidth}\raggedright
\textbf{Notation}
\end{minipage} & \begin{minipage}[b]{\linewidth}\raggedright
\textbf{Definition}
\end{minipage} \\
\midrule\noalign{}
\endhead
\bottomrule\noalign{}
\endlastfoot
\(m\) & Number of active elements. \\
\(A\) & Block of all \(m\) active elements to be sorted. \\
\(B\) & Block of \(3m - 1\) buffer elements, identifiable from active
elements by a single comparison. \\
\(V\) & Bit vector of \(m/\lg^{2}m\) bits. reading requires \(1\)
comparison, writing \(3\) moves. \\
\end{longtable}
}

\textbf{Table E.2:} \hyperref[structure-of-the-memory]{Section 2.1} Notation
Table

{\def\LTcaptype{none} 
\begin{longtable}[]{@{}
  >{\raggedright\arraybackslash}p{(\linewidth - 2\tabcolsep) * \real{0.1246}}
  >{\raggedright\arraybackslash}p{(\linewidth - 2\tabcolsep) * \real{0.8754}}@{}}
\toprule\noalign{}
\begin{minipage}[b]{\linewidth}\raggedright
\textbf{Notation}
\end{minipage} & \begin{minipage}[b]{\linewidth}\raggedright
\textbf{Definition}
\end{minipage} \\
\midrule\noalign{}
\endhead
\bottomrule\noalign{}
\endlastfoot
\(F\) & Frame memory, part of \(B\) containing frame elements. \\
\(G\) & Cache memory, part of \(B\) made up of cache lines. \\
\(S\) & Segment memory, part of \(B\) made up of segments. \\
\(r_{\#}\) & Number of frame elements/cache lines/segment pointers/cache
integers. \\
\(s_{\#}\) & Number of additional segments, i.e. segments exluding $S_0$. \\
\(f_{i}\) & The \(i\)th frame element in \(F\). \\
\(f_{i}'\) & The \(i\)th active frame element in \(F\). \\
\(G_{i}\) & The \(i\)th cache line in \(G\), corresponding with
\(f_{i}\). \\
\(G_{i}'\) & The \(i\)th active cache line in \(G\), corresponding with
\(f_{i}'\). \\
\(g\) & Max capacity of a cache line. \\
\(S_{i}\) & The \(i\)th segment in \(S\), corresponding with
\(f_{i}'\). \\
\(s\) & Length of a segment. \\
\(P\) & Pointer data in \(V\). \\
\(P_{i}\) & The \(i\)th segment pointer in \(P\), corresponding with
\(f_{i}\). \\
\(P_{i}'\) & The \(i\)th active segment pointer in \(P\),
corresponding with \(f_{i}'\). \\
\(I_i\) & The \(i\)th cache integer in \(I\), corresponding with
\(G_{i}\). \\
\(I_i'\) & The \(i\)th active cache integer in \(I\),
corresponding with \(G_{i}'\). \\
\(R_{i}\) & The \(i\)th representative, a compound data type
formed by \(f_{i}\), \(G_{i}\), \(P_{i}\) and \(I_i\). \\
\(R_{i}'\) & The \(i\)th active representative, corresponding with
\(S_{i}\). Defined as \(R_{j}\) where \(j\) is the \(i\)th smallest
value such that \(f_{j}\) is active. \\
\end{longtable}
}

\textbf{Table E.3:} \hyperref[structure-of-segment-representatives]{Section 2.2}
and
\hyperref[appendix-a-detailed-procedures-and-proofs-for-section-2.2]{App.
A} Notation Table

{\def\LTcaptype{none} 
\begin{longtable}[]{@{}
  >{\raggedright\arraybackslash}p{(\linewidth - 2\tabcolsep) * \real{0.1346}}
  >{\raggedright\arraybackslash}p{(\linewidth - 2\tabcolsep) * \real{0.8654}}@{}}
\toprule\noalign{}
\begin{minipage}[b]{\linewidth}\raggedright
\textbf{Notation}
\end{minipage} & \begin{minipage}[b]{\linewidth}\raggedright
\textbf{Definition}
\end{minipage} \\
\midrule\noalign{}
\endhead
\endlastfoot
\(r_{\#}\) & Number of representatives. \\
\(s_{\#}\) & Number of active representatives, equivalent to the definition in Table E.2. \\
\(R\) & Block of all \(r_{\#}\) representatives. \\
\(R_{i}\) & The \(i\)th representatives. \\
\(R_{\textbf{i}}'\) & The \(i\)th active representative. \\
\(L\) & Library zone, containing sorted sequence of all active
representatives \(R_{1}'\ldots R_{s_{\#}}'\) interlaced with buffer
positions. \\
\(L_{i}\) & The \(i\)th gap in \(L\); a contiguous block of \(8\)
representatives \(\mathcal{\ell}_{1}\ldots\mathcal{\ell}_{8}\). \\
\(s_{\#}'\) & Number of gaps in \(L\). \\
\(L_{1}\ldots L_{s_{\#}'}\) & Gaps in the library, which are contiguous
blocks of \(8\) representatives
\(\mathcal{\ell}_{1}\ldots\mathcal{\ell}_{8}\). \\
\(D\) & Rebuild buffer. \\
\(n_{1}\ldots n_{2s_{\#}' - 1}\) & Logical segment tree over gaps
\(L_{1}\ldots L_{s_{\#}}\). \\
\(\text{size}(n_{i})\) & Number of active representatives in gaps covered by
node \(n_{i}\). \\
\(\text{span}(n_{i})\) & Number of gaps covered by node \(n_{i}\). \\
\(\psi(n_{i})\) & Imbalance of node \(n_{i}\), used for amortized
analysis. \\
\end{longtable}
}

\textbf{Table E.4:} \hyperref[sorting-small-subsequences]{Section 2.3} and
\hyperref[appendix-b-full-small-sort-procedure]{App. B} Notation Table

{\def\LTcaptype{none} 
\begin{longtable}[]{@{}
  >{\raggedright\arraybackslash}p{(\linewidth - 2\tabcolsep) * \real{0.1346}}
  >{\raggedright\arraybackslash}p{(\linewidth - 2\tabcolsep) * \real{0.8654}}@{}}
\toprule\noalign{}
\begin{minipage}[b]{\linewidth}\raggedright
\textbf{Notation}
\end{minipage} & \begin{minipage}[b]{\linewidth}\raggedright
\textbf{Definition}
\end{minipage} \\
\midrule\noalign{}
\endhead
\bottomrule\noalign{}
\endlastfoot
\(m'\) & size of small subsequence to be sorted. \\
\(w\) & size of each word. \\
\(A'\) & Small subsequence of \(m'\) active elements. \\
\(B'\) & Merge buffer of \(m'\) buffer elements. \\
\(k\) & size of runs sorted by table sort/number of runs in a group
merged at once. \\
\(k_{\#}\) & Number of merge passes to sort \(A'\). \\
\(r_{i}\) & size of runs after the \(i\)th merge pass (excluding the
last run that is potentially short). \\
\(A_{i}\) & The \(i\)th sorted run from the previous merge pass. \\
\(B_{i}\) & The \(i\)th block of buffer elements reserved for merging
the \(i\)th group of \(k\) runs. \\
\(k'\) & Number of runs merged in the final merge pass. \\
\end{longtable}
}

\textbf{Table E.5:} \hyperref[structure-of-cache-lines]{Section 2.6} and
\hyperref[appendix-c-analysis-of-cache-lines]{App. C} Notation Table

{\def\LTcaptype{none} 
\begin{longtable}[]{@{}
  >{\raggedright\arraybackslash}p{(\linewidth - 2\tabcolsep) * \real{0.1346}}
  >{\raggedright\arraybackslash}p{(\linewidth - 2\tabcolsep) * \real{0.8654}}@{}}
\toprule\noalign{}
\begin{minipage}[b]{\linewidth}\raggedright
\textbf{Notation}
\end{minipage} & \begin{minipage}[b]{\linewidth}\raggedright
\textbf{Definition}
\end{minipage} \\
\midrule\noalign{}
\endhead
\bottomrule\noalign{}
\endlastfoot
\multicolumn{2}{@{}>{\raggedright\arraybackslash}p{(\linewidth - 2\tabcolsep) * \real{1.0000} + 2\tabcolsep}@{}}{%
Randomized Approach} \\
\(g\) & Max capacity of a cache line. \\
\(g_{i}\) & The \(i\)th element in a cache line. \\
\(d\) & The cache integer to store integers from \(0\) to \(g\) made up
of bits \(d_{0}\ldots d_{\left\lfloor \lg g \right\rfloor}\) where
\(d_{j}\) has priority \(2^{j}\). \\
\multicolumn{2}{@{}>{\raggedright\arraybackslash}p{(\linewidth - 2\tabcolsep) * \real{1.0000} + 2\tabcolsep}@{}}{%
Deterministic Approach} \\
\(t\) & Parameter for comparisons moves tradeoff, obeys
\(2 \leq t \leq \lg^{*}n - 1\). \\
\(t'\) & Number of levels in a cache line, follows \(t' \leq t - 1\) by
definition. \\
\(G_{i}\) & The cache block at level \(i\). \\
\(g_{i}\) & size of the cache block at the \(i\)th level. \\
\(g\) & Max capacity of a cache line. \\
\(A_{i}\) & The \(i\)th sub cache block of active elements. \\
\(B_{i}\) & The \(i\)th sub cache block of buffer elements. \\
\end{longtable}
}

\textbf{Table E.6:} \hyperref[sorting-in-place]{Section 3} and
\hyperref[appendix-d-in-place-conversion]{App. D} Notation Table

{\def\LTcaptype{none} 
\begin{longtable}[]{@{}
  >{\raggedright\arraybackslash}p{(\linewidth - 2\tabcolsep) * \real{0.1346}}
  >{\raggedright\arraybackslash}p{(\linewidth - 2\tabcolsep) * \real{0.8654}}@{}}
\toprule\noalign{}
\begin{minipage}[b]{\linewidth}\raggedright
\textbf{Notation}
\end{minipage} & \begin{minipage}[b]{\linewidth}\raggedright
\textbf{Definition}
\end{minipage} \\
\midrule\noalign{}
\endhead
\bottomrule\noalign{}
\endlastfoot
\(n\) & Number of elements in the input array. \\
\(\mathcal{A}\) & Input array of \(n\) elements. \\
\(L\) & Block containing the smallest \(n/\lg^{2}n\) elements used to
simulate a bit vector with \(R\). \\
\(R\) & Block containing the largest \(n/\lg^{2}n\) elements used to
simulate a bit vector with \(L\). \\
\(\mathcal{A'}\) & The remaining portion of \(\mathcal{A}\) excluding
\(L\) and \(R\). \\
\(\mathcal{A}_{S}\) & Section of \(\mathcal{A'}\) that is already in its
final sorted position. \\
\(\mathcal{A}_{U}\) & Unsorted portion of \(\mathcal{A'}\). \\
\(n_{i}\) & size of \(\mathcal{A}_{U}\) at the \(i\)th iteration. \\
\(p\) & Buffer separator to differentiate active and buffer elements. \\
\(A_{<}\) & Block containing active elements \(< p\). \\
\(n_{i, <}\) & Number of active elements at the \(i\)th iteration, i.e.
\(|A_{<}|\). \\
\(B_{\geq}\) & Block containing buffer elements \(\geq p\). \\
\(n_{i, \geq}\) & Number of buffer elements at the \(i\)th iteration,
i.e. \(|B_{\geq}|\). \\
\(A_{< ,S}\) & Sorted permutation of block \(A_{<}\). \\
\(B_{\geq}'\) & Permutation of block \(B_{\geq}\) after sorting
\(A_{<}\). \\
\(P\) & Block of elements \(= p\). \\
\(n_{i, =}\) & Number of elements \(= p\) at the \(i\)th iteration, i.e.
\(|P|\). \\
\(B_{>}\) & Block of elements \textgreater{}\(p\). \\
\(I\) & Number of iterations to sort the entire input array
\(\mathcal{A}\). \\
\end{longtable}
}

\end{center}

\end{document}